\documentclass[12pt]{article}
\pdfoutput=1
\usepackage{amsmath,amssymb}
\usepackage{epsfig}
\usepackage{cite}

\def\m@th{\mathsurround=0pt }
\def\leftrightarrowfill{$\m@th \mathord\leftarrow \mkern-6mu
	\cleaders\hbox{$\mkern-2mu \mathord- \mkern-2mu$}\hfill
	\mkern-6mu \mathord\rightarrow$}

\def\overleftrightarrow#1{\vbox{\ialign{##\crcr
	\leftrightarrowfill\crcr\noalign{\kern-1pt\nointerlineskip}
	$\hfil\displaystyle{#1}\hfil$\crcr}}}

\newcommand{\be}{\begin{equation}}
\newcommand{\ee}{\end{equation}}

\newcommand{\Tr}{\mathop{\rm Tr}}
\def\I{\rm 1\kern-.24em l}  

\def\shat{\ifmmode \hat{s}\else $\hat{s}$\fi}

\def\UV{UV}
\def\IR{IR}
\def\nn{\nonumber}

\newcommand{\newc}{\newcommand}

\newc{\gsim}{\lower.7ex\hbox{$\;\stackrel{\textstyle>}{\sim}\;$}}
\newc{\lsim}{\lower.7ex\hbox{$\;\stackrel{\textstyle<}{\sim}\;$}}
\newc{\ie}{{\it i.e.}}
\newc{\etal}{{\it et al.}}
\newc{\mev}{\hbox{\rm\,MeV}}
\newc{\gev}{\hbox{\rm\,GeV}}
\newc{\tev}{\hbox{\rm\,TeV}}
\newc{\xpb}{\hbox{\rm\, pb}}
\newc{\xfb}{\hbox{\rm\, fb}}

\newc{\G}{{\cal G}}
\newc{\h}{{\cal H}}
\newc{\D}{{\cal D}}
\newc{\E}{{\cal E}}
\newc{\x}{{\widehat x}}
\newc{\q}{{\widehat q}}

\newc{\mtop}{m_t}
\newc{\mbot}{m_b}
\newc{\mz}{M_Z}
\newc{\mw}{M_W}
\newc{\alphasmz}{\alpha_s(M_Z)}
\newc{\swsq}{\sin^2\theta_W}
\newc{\cwsq}{\cos^2\theta_W}
\newc{\tw}{\tan\theta_W}
\newc{\cw}{\cos\theta_W}
\newc{\sw}{\sin\theta_W}
\newc{\BR}{\hbox{\rm BR}}
\newc{\zbb}{Z\to b\bar}
\newc{\Gb}{\Gamma (Z\to b\bar b)}
\newc{\Gh}{\Gamma (Z\to \hbox{\rm hadrons})}
\newc{\sgn}{\mbox{sgn}}

\def\I{1\hspace{-4pt}1}
\def\ov{\overline}

\newcounter{mysubequation}[equation]

\def\beq{\begin{equation}}
\def\eeq{\end{equation}}
\def\bea{\begin{eqnarray}}
\def\eea{\end{eqnarray}}
\def\slashchar#1{\setbox0=\hbox{$#1$}           
   \dimen0=\wd0                                 
   \setbox1=\hbox{/} \dimen1=\wd1               
   \ifdim\dimen0>\dimen1                        
      \rlap{\hbox to \dimen0{\hfil/\hfil}}      
      #1                                        
   \else                                        
      \rlap{\hbox to \dimen1{\hfil$#1$\hfil}}   
      /                                         
   \fi}                                         %
\catcode`@=11
\long\def\@caption#1[#2]#3{\par\addcontentsline{\csname
  ext@#1\endcsname}{#1}{\protect\numberline{\csname
  the#1\endcsname}{\ignorespaces #2}}\begingroup
    \small
    \@parboxrestore
    \@makecaption{\csname fnum@#1\endcsname}{\ignorespaces #3}\par
  \endgroup}
\catcode`@=12

\def\UV{\rm UV}
\def\IR{\rm IR}
\def\Le{{\bf L}}
\def\R{{\bf R}}

\begin{document}

\baselineskip=18pt

\setcounter{footnote}{0}
\setcounter{figure}{0}
\setcounter{table}{0}

\begin{titlepage}
\begin{flushright}
\end{flushright}
\vspace{.3in}

\begin{center}
{\Large \bf
Nucleon Form Factors from 5D Skyrmions
}

\vspace{0.5cm}

{\bf Giuliano Panico$^{a}$ and Andrea Wulzer$^{b}$}

\vspace{.5cm}

\centerline{$^{a}${\it  Bethe Center for Theoretical Physics and
Physikalisches Institut der Universit\"at Bonn,}}
\centerline{\it Nussallee 12, 53115 Bonn, Germany}
\centerline{$^{b}${\it Institut de Th\'eorie des Ph\'enom\`enes Physiques, EPFL,  CH--1015 Lausanne, Switzerland}}

\end{center}
\vspace{.8cm}

\begin{abstract}
\medskip
\noindent
Several aspects of hadron physics are well described by a simple 5D effective field theory. Baryons arise in this
scenario as ``large'' (and therefore calculable) 5D skyrmions. We extend and refine the existing analysis of this
5D soliton, which is fairly non-trivial due to the need of numerical methods. We perform the complete quantization
of those collective coordinates which are relevant for computing the static observables like the nucleon form factors.
We compare the result with simple expectations about large-$N_c$ QCD and with the experimental data.
An agreement within $30\%$ is found.
\end{abstract}

\bigskip
\bigskip

\end{titlepage}


\section{Introduction and Conclusions}

Certain 5D effective gauge theories, often referred to as ``Holographic QCD'' or
``$\rm{AdS}/\rm{QCD}$" models
\cite{Son:2003et,Erlich:2005qh,DaRold:2005zs,DaRold:2005vr}, closely
resemble low-energy QCD in the limit of large number of colors $N_c$.
The similarity is qualitative as
these theories contain, like large-$N_c$ QCD, infinite towers of weakly interacting
mesons, but also quantitative. Leading order calculations in such 5D models typically
describe the physics of the lightest mesons to $10\%$ accuracy in terms of an extremely
limited number of parameters.
These results are compatible with the hypothesis that leading order calculations in the 5D
model reproduce large-$N_c$ QCD.

Baryons arise in this scenario as solitons with a conserved topological charge which represents
 the baryon number. They are the 5D analog of skyrmions
\cite{Skyrme:1961vq,Adkins:1983ya} (see \cite{Meissner:1987ge}  for a review), so
 we will refer to them as 5D skyrmions. They differ from ordinary skyrmions,
however, in the important aspect of calculability, as shown in
\cite{Pomarol:2007kr,Pomarol:2008aa}. The skyrmion solutions obtained in
4D models of mesons --which may or may not contain some vector resonances on top
of the pion field-- have a size which is of the order of the inverse cut-off of the theory,
and incalculable UV effects do not decouple. This problem, which constitutes the main
reason of theoretical dissatisfaction about the 4D Skyrme model, is solved in the 5D case
because the size of the 5D skyrmion is larger than the 5D cut-off.

A string construction named ``Sakai--Sugimoto model" \cite{Sakai:2004cn}
might provide a UV completion of the $\rm{AdS}/\rm{QCD}$ models and give them an
interpretation in the standard framework of $\rm{AdS}/\rm{CFT}$. In the low-energy
supergravity limit of large 't Hooft coupling $\lambda\rightarrow\infty$ the
Sakai--Sugimoto model reduces indeed to a 5D theory with $U(N_f)$ gauge
symmetry ($N_f$ denotes the number of flavors)
and two $\rm{AdS}_5$-like boundaries on which the sources for Left- and
Right-handed currents are located. This can be rewritten as a
$U(N_f)_L\times U(N_f)_R$ theory living on one-half of the space with one
$\rm{AdS}_5$-like (UV) boundary  on which both sources live and one IR
boundary on which symmetry-breaking conditions as
in eq.~(\ref{irboundary condition}) are imposed. The Sakai--Sugimoto model
is almost equivalent, in the limit in which practical calculations are performed,
to the effective theory considered in the present paper. There is however a difference
which, as remarked in \cite{Pomarol:2007kr,Pomarol:2008aa}, becomes extremely
relevant in the baryon sector. In the Sakai--Sugimoto model the effective 5D interaction scale
$M_5$ is proportional to the 't Hooft coupling $\lambda$, {\it{i.e.}} $M_5\propto \lambda$,
while the coefficient of the Chern--Simons (CS) term, which is fixed by the
Adler--Bardeen anomaly, has no $\lambda$ factor. This implies, given the definition
in eq.~(\ref{gamma}), $\gamma\propto1/\lambda\rightarrow0$. The parameter
$\gamma$ controls the size of the skyrmion, $\rho\propto\gamma^{1/2}$ in the
Sakai--sugimoto model and for this
reason the string effects, which are encoded in higher dimensional operators, do not
decouple and there is no advantage with respect to the usual 4D Skyrme model
in what concerns calculability. It is interesting,
nevertheless, to forget about higher dimensional operators and study baryons in
this framework, describing them in terms of ``small" Yang--Mills instantons
\cite{Hata:2007mb} or in terms of 4D Skyrmions \cite{Nawa:2006gv}.

In the present paper we complete and refine the analysis of the 5D skyrmions
presented in \cite{Pomarol:2008aa}, with the aim of computing nucleon static
observables and in particular the current form factors. To this end we need
to perform a complete quantization of the 5D skyrmion collective coordinates,
which is a non-trivial task as it requires to solve numerically a new set of
partial differential equations. Obtaining predictions for the form
factors at non-zero transfer momentum requires, moreover, an increased
precision of the solution, which we obtain by refining our numerical method.
The nucleon form factors
in the Sakai--Sugimoto model have been computed in  \cite{Hashimoto:2008zw}
 (see also \cite{Hong:2007dq,Kim:2008pw}),
by performing a ``small-size'' ({\it{i.e.}} small $\gamma$) expansion in
which analytical results can be obtained. \footnote{The same calculation has  been performed in
\cite{Hata:2008xc} with a different (and erroneous, in our understanding) definition of the chiral currents.}
This expansion is not trustable in our case because, as explained above, the
size of the 5D skyrmions is large and $\gamma\sim1$.

The paper is organized as follows. In sect.~2, after a brief review of the model and
of the static
5D skyrmion solution we identify the zero-mode fluctuations which
are relevant to describe static properties and we discuss the corresponding
collective coordinates classical Lagrangian. A suitable ansatz is described which
permits to rewrite in a 2D form the 4D equations which define the zero-modes.
Sect.~3 is devoted to the collective coordinate quantization and to the calculation
of the form factors, this discussion is basically the same as in the 4D Skyrme model
\cite{Adkins:1983ya,Meissner:1987ge}, though adapted to the present case.
Sect.~4 contains a detailed presentation of our results.
After the comparison with
simple expectations about large-$N_c$ QCD we discuss the
divergences due to the chiral limit and we check that the Goldberger--Treiman
relation holds in our model. Finally, we compare our results with experimental data
and find a level of agreement better than
 $30\%$ for all the observables, with the
notable exception of the axial coupling $g_A$ for which we find $g_A=0.70$
versus an experimental value $g_A=1.25$.\footnote{An erroneous value of $g_A$
was reported in \cite{Pomarol:2008aa}. The error was due to a subtlety, which we
will discuss in the following, in taking the zero momentum limit of the axial form
factor, combined with a more trivial mistake.}
Most of the technical details are presented
in the appendices. In appendix~A the 2D equations of motions and boundary
conditions are derived while appendix~B gives some detail on the numerical
techniques we employed to obtain the solution.

In spite of the failure in the axial coupling, the level of accuracy of our results is consistent
with the expected size of the $1/N_c$ corrections or, which is the same, the expected
size of next-to-leading contributions in our model. It is not unreasonable that anomalously
large numerical factors could change into $80\%$ the naively expected $30\%$ correction
to $g_A$. Such large $1/N_c$ corrections arise for example if one follows, instead of 
the approach we consider, the quantization procedure of the collective coordinates 
proposed in \cite{Amado:1986ef}. This ``alternative'' quantization is equivalent to 
the standard one at the leading order in $1/N_c$ but it also contains large $1/N_c$ 
corrections. We will discuss in sect.~4 how these corrections change our predictions. 
In the case of $g_A$ we find, remarkably, the much better result $g_A=1.17$ while the level 
of the agreement of the other observables is unaffected. 
\footnote{We thank the referee of Nucl.~Phys.~A for suggesting this possibility to us.}
Even without applying this correction, the results which we obtain 
are significantly more accurate than those of the
original Skyrme model \cite{Adkins:1983ya} (in which, we remark, $g_A$ is also small,
 $g_A=0.65$),
but not as good as those of more refined skyrmion models
(which, of course, also have more parameters) like the ones reviewed in
\cite{Meissner:1987ge}. It seems, as we will discuss in sect.~4, that the inclusion of the
explicit breaking of the chiral symmetry ({\it{i.e.}} of the pion mass $m_\pi$)
will improve the
agreement of several observables and that $g_A$ may display an enhanced
sensitivity to $m_\pi$. It is certainly worth exploring this direction.

\section{Skyrmions in 5D}

\subsection{The Model}

We will consider the same model as in \cite{Pomarol:2008aa}, {\it i.e.} a $U(2)_L\times U(2)_R$ gauge theory in five
dimensions with metric $ds^2=a(z)^2\left(\eta_{\mu\nu}dx^\mu dx^\nu-dz^2\right)$, where we denoted as $x^\mu$
the usual $4$ coordinates with mostly minus metric and with $z$, which runs in the interval $[z_{\UV},z_{\IR}]$,
the extra dimension. We choose our metric to be AdS$_5$ and therefore the warp factor $a(z)$ to be
\begin{equation}
a(z)=\frac{z_{\IR}}{z}\, ,
\label{warpf}
\end{equation}
with $z_{\UV}\rightarrow 0$ to be taken at the end of the calculations. In this limit, $z_{\IR}$ coincides with the
conformal length $L=\int^{z_{\IR}}_{z_{\UV}} dz=z_{\IR}-z_{\UV}$. It should be kept in mind that, since gravity is
non-dynamical in our model, the choice of the warp factor $a(z)$ is arbitrary. It is commonly believed, however, that
different ``reasonable'' choices of $a(z)$ would not affect in a significant way the predictions for IR observables,
like those we will compute in this paper.\footnote{This belief is supported by Ref.~\cite{Hirn:2005nr}, in which IR
predictions for flat and AdS$_5$ spaces were compared, and by Ref.~\cite{DaRold:2005zs},
in which departures from AdS$_5$ were considered. Moreover, the Sakai--Sugimoto model is equivalent, for what
calculations in the meson sector are concerned,  to a 5D model of the kind we are considering with non-AdS$_5$ warped
metric. The predictions of this model are very similar to those of $\rm{AdS}/\rm{QCD}$, again suggesting that the choice of the
metric is not so relevant.} Choosing AdS$_5$ --or at least a geometry with an ``AdS$_5$-like'' boundary--
is crucial, on the contrary, if one wants to match UV correlation functions with those
computed in QCD by perturbation theory
\cite{Son:2003et,DaRold:2005zs,DaRold:2005vr,Erlich:2005qh,Hirn:2005nr}.
For this reason, in the literature the choice in eq.~(\ref{warpf}) is commonly adopted.

We will denote the $U(2)_L$ and $U(2)_R$ gauge connections respectively by $\Le_M$ and $\R_M$, where $M=\{\mu,5\}$, and
parametrize them as $\Le_M=L_M^a\sigma_a/2+\widehat{L}_M\I/2$ and $\R_M=R_M^a\sigma_a/2+\widehat{R}_M\I/2$ in terms of
the Pauli matrices $\sigma_a$ and the identity $\I$. Chiral symmetry is broken at the $z=z_{\IR}$ boundary
(IR-boundary) by the following conditions:
\beq
\left(\Le_\mu-\R_\mu\right)\left|_{z=z_{\IR}}\right.=0\ ,\;\;\;\;\;\;\left(\Le_{\mu 5}+\R_{\mu 5}\right)\left|_{z=z_{\IR}}\right.=0\, ,
\label{irboundary condition}
\eeq
where the 5D field strength is defined as $\Le_{MN}=\partial_M \Le_N-\partial_N \Le_M-i[\Le_M,\,\Le_N]$, and analogously for $\R_{MN}$. On the other boundary, the UV one, we impose Dirichlet conditions:
\beq
\Le_\mu\left|_{z=z_{\UV}}\right.=\,0\ , \;\;\;\;\;\; \R_\mu\left|_{z=z_{\UV}}\right.=\,0
\, .
\label{uvboundary condition}
\eeq

The 5D action $S=S_g+S_{CS}$ consists of a standard gauge kinetic part
\beq
S_g=-\int d^4{x}\int^{z_{\IR}}_{z_{\UV}} dz\,  a(z)\, \frac{M_5}{2} \left\{
\Tr\left[{L_{MN}L^{MN}}\right]\,+\,\frac{\alpha^2}2{\widehat{L}}_{MN}{\widehat{L}}^{MN}\,+\,\{L\,\leftrightarrow\,R\}\right\}\, ,
\label{Sg}
\eeq
and of a Chern--Simons part
\be
S_{CS}\,=\,\frac{N_c}{16\pi^2}\int d^5x\left\{
\frac14\epsilon^{MNOPQ}{\widehat{L}_M}\Tr\left[L_{NO}L_{PQ}\right]\,+\,
\frac1{24}\epsilon^{MNOPQ}{\widehat{L}_M}{\widehat{L}_{NO}}{\widehat{L}_{PQ}}\,-\,\{L\,\leftrightarrow\,R\}
\right\}\, .
\label{Scs}
\ee
The $S_{CS}$ is needed to reproduce the QCD anomalies and its coefficient is fixed to be proportional to the number
of colors $N_c$.

In order to compare our 5D model with the real world, and in particular to compute the form factors as we will do in this paper, we need to identify the chiral currents to which the electroweak bosons are coupled.
These operators, which would be given in QCD by the quark bilinears $j_{\mu,L\,(R)}^a={\overline{Q}}_{L\,(R)}\gamma^\mu \sigma^a/2 Q_{L\,(R)}$,
${\widehat{j}}_{\mu,L\,(R)}={\overline{Q}}_{L\,(R)}\gamma^\mu \I/2 Q_{L\,(R)}$
 correspond in our model to \cite{Pomarol:2008aa}
\be
J_{L\, \mu}^{a}\,=\,M_5\big(a(z)L_{\mu\,5}^a\big)\left|_{z=z_{\UV}}\right.\ ,
\;\;\;\;\;{\widehat J}_{L\, \mu}\,=\,\alpha^2M_5\big(a(z){\widehat L}_{\mu\,5}\big)\left|_{z=z_{\UV}}\right.\, ,
\label{cur0}
\ee
and analogously for $R$.

It is important to remark that this model, as discussed in \cite{Pomarol:2008aa}, is a valid effective field theory
with an NDA cut-off $\Lambda_5$ which is bigger than the scale of the lightest resonances
and can be sent to infinity for $M_5\rightarrow\infty$.
This allows us to include
only the lowest dimensional operators in the action (\ref{Sg},\ref{Scs})
since the others, which would surely arise in a
UV completion of the model, are expected to give a subleading contribution. At the leading order this model is
extremely predictive: its only 3 parameters are $M_5$, $L$ and $\alpha$.
The 5D interaction scale $M_5$ can be traded
for the adimensional parameter
\be
\gamma=\frac{N_c}{16\pi^2 M_5 L\alpha}\,,
\label{gamma}
\ee
which controls the size of the skyrmion; $\rho\sim\gamma^{2/3}$.
We want to interpret the 5D weak coupling expansion as
the $1/N_c$ expansion, therefore we will take the interaction scale to scale
like $N_c$, {\it i.e.} $M_5\propto N_c$ so that $\gamma,\rho\propto N_c^0$.

\subsection{The Static Soliton Solution}
\label{ss}

Our model admits topologically non-trivial static solutions of the classical equations of motions (EOM).
These are identified with the baryons and therefore the topological charge
\be
B=\frac1{32\pi^2}\int d^3x\int^{z_{\IR}}_{z_{\UV}} dz\,
\epsilon_{\hat\mu\hat\nu\hat\rho\hat\sigma}\Tr\left[{L^{\hat\mu\hat\nu}L^{\hat\rho\hat\sigma}}
-{R^{\hat\mu\hat\nu}R^{\hat\rho\hat\sigma}}\right]\,,
\label{Bch}
\ee
is identified with the baryon number. The indices $\hat\mu,\,\hat\nu,\,\ldots$ label,
throughout the paper, the 4 spatial
coordinates, but they are raised with Euclidean metric.

Regular static solutions with $B=1$ have been found in \cite{Pomarol:2008aa}. The non-vanishing components of
the $R$ fields can be written in terms of 2D fields as
\be
\left\{
\begin{array}{l}
\displaystyle
{\ov R}^a_j({\bf x},z) = \displaystyle A_1(r,z) \x_a \x_j + \frac1r \varepsilon_{ajk} \x_k
-\frac{\phi_{(x)}}r\varepsilon^{(x,y)}\Delta^{(y),aj} \,,\\
\displaystyle
{\ov R}^a_5({\bf x},z) = \displaystyle  A_2(r,z)\x^a \,,\\
\displaystyle
\alpha\widehat {\ov R}_0({\bf x},z) = \displaystyle \frac{s(r,z)}r \,,
\end{array}
\right.
\label{sts}
\ee
where $r^2=\sum_i x^i x^i$, \  $\x^i=x^i/r$, $\varepsilon^{(x,y)}$ is the antisymmetric tensor with
$\varepsilon^{(1,2)}=1$ and the ``doublet'' tensors $\Delta^{(1,2)}$ are
\be
\Delta^{(x),ab}\,=\, \left[
\begin{array}{l}
 \epsilon^{abc}\x^c\\
\x^a\x^b - \delta^{ab}
\end{array}
\right]\,.
\label{defde}
\ee
Due to parity invariance \mbox{$\{ L\leftrightarrow R, {\bf  x}\leftrightarrow -{\bf x}\}$} we restrict, in both the
static and non--static case which we will consider in the next section, to configurations for which
$L_i({\bf x},z,t)=-R_i(-{\bf x},z,t)$, $L_{5,0}({\bf  x},z,t)=R_{5,0}(-{\bf x},z,t)$ and analogously
for ${\hat L}$, ${\hat  R}$. Eq.~(\ref{sts}) therefore defines the static solution completely.

It is important to remark that the static solution in eq.~(\ref{sts}) is ``cylindrically'' symmetric, meaning that
it is invariant under the simultaneous action of 3D space rotations $x_a\sigma^a\,\rightarrow\,r^\dagger x_a\sigma^ar$,
with $r\in SU(2)$, and vector $SU(2)$ global transformations $L,R\rightarrow r\,(L,R)\,r^\dagger$. An equivalent way to
state this is that a 3D rotation with $r$ acts on the solution (\ref{sts}) exactly as an $SU(2)$ vector one in the
opposite direction ({\it i.e.} with $r^\dagger$) would do.

\subsection{Zero-Mode Fluctuations}

Let us now consider time-dependent infinitesimal deformations of the static solutions. Among these, the zero-mode
({\it i.e.} zero frequency) fluctuations are particularly important as they will describe single-baryon states.
Zero-modes can be defined as directions in the field space in which uniform and slow motion is permitted by the
classical dynamics and they are associated with the global symmetries of the problem, which are in our case $U(2)_V$
and $3$-space rotations plus $3$-space translations. The latter would describe baryons moving with uniform velocity
and therefore can be ignored in the computation of static properties like the form factors. Of course, the global
$U(1)_V$ acts trivially on all our fields and the global $SU(2)_V$ has the same effect as $3$-space rotations on
the static solution (\ref{sts}) because of the cylindrical symmetry. The space of static solutions which are
of interest for us is therefore parametrized by $3$ real coordinates --denoted as collective coordinates--
which define an $SU(2)$ matrix $U$.

To construct zero-modes fluctuations we consider collective coordinates with general time dependence, {\it i.e.}
we perform a global $SU(2)_V$ transformation on the static solution
\be
R_{\hat\mu}({\bf x},z;U)\,=\,U\,{\ov R}_{\hat\mu}({\bf x},z)\,U^\dagger\  ,
\;\;\;\;\;
{\widehat{R}_0}({\bf x},z;U)\,=\,{{{\widehat{\ov{R}}}}}_0({\bf x},z)\,,
\label{eq0m}
\ee
but we allow $U=U(t)$ to depend on time. It is only for constant $U$ that eq.~(\ref{eq0m}) is a solution of
the time-dependent EOM. For infinitesimal but non-zero rotational velocity
$$
K=k_a\sigma^a/2=-i U^\dagger dU/dt\,,
$$
eq.~(\ref{eq0m}) becomes an infinitesimal deformation of the static solution. Along the zero-mode direction
uniform and slow motion is classically allowed, for this reason our fluctuations should fulfill the
time-dependent EOM at linear order in $K$ provided that $d K/dt=0$.

From the action (\ref{Sg},\ref{Scs}) the following EOM are derived
\be
\left\{
\begin{array}{l}
\displaystyle
D_{\hat\nu}\left(a(z)R^{\hat\nu}_{\;0}\right)+\frac{\gamma  \alpha L}4\epsilon^{\hat\nu\hat\omega\hat\rho\hat\sigma}R_{\hat\nu\hat\omega}{\widehat R}_{\hat\rho\hat\sigma}=0  \\
\displaystyle
\alpha\partial_{\hat\nu}\left(a(z){\hat R}^{\hat\nu}_{\;\  0}\right)+\frac{\gamma  L}4\epsilon^{\hat\nu\hat\omega\hat\rho\hat\sigma}\left[{\rm  Tr}\left(R_{\hat\nu\hat\omega}R_{\hat\rho\hat\sigma}\right)  +\frac12{\widehat R}_{\hat\nu\hat\omega}{\widehat  R}_{\hat\rho\hat\sigma}\right]=0\\
\displaystyle
D_{\hat\nu}\left(a(z)R^{\hat\nu\hat\mu}\right)-a(z)D_0R_{0}^{\;\  \hat\mu}-\frac{\gamma\alpha   L}2\epsilon^{\hat\mu\hat\nu\hat\rho\hat\sigma}\left[R_{\hat\nu  0}{\widehat R}_{\hat\rho\hat\sigma}+R_{\hat\nu \hat\rho}{\widehat  R}_{\hat\sigma 0}\right]=0\\
\displaystyle
\alpha\partial_{\hat\nu}\left(a(z){\widehat  R}^{\hat\nu\hat\mu}\right)-\alpha a(z)\partial_0{\widehat R}_{0}^{\;\  \hat\mu}-\gamma L\epsilon^{\hat\mu\hat\nu\hat\rho\hat\sigma}\left[{\rm  Tr}\left(R_{\hat\nu 0} R_{\hat\rho\hat\sigma}\right)+\frac12{\widehat  R}_{\hat\nu 0}{\widehat R}_{\hat\rho\hat\sigma}\right]=0
\end{array}
\right. \, .
 \label{eomt}
 \ee
We only need to specify the EOM for one chirality since we are considering, as explained in the previous section,
a parity invariant ansatz. We would like to find solutions of eq.~(\ref{eomt}) for which  $R_{\hat\mu}$ and
${\widehat R}_{0}$ are of the form (\ref{eq0m}); it is easy to see that the  time-dependence of $U$ in
eq.~(\ref{eq0m}) acts as a source for the components $R_{0}$ and ${\widehat  R}_{\hat\mu}$, which therefore cannot
be put to zero as in the static case. Notice that the same happens in the case of the 4D  skyrmion
\cite{Meissner:1987ge}, in which  the temporal and spatial components of the $\rho$ and $\omega$ mesons are turned
on in the rotating skyrmion solution. Also, it can be shown that eq.~(\ref{eomt}) can be solved, to linear order
in $K$ and for $dK/dt=0$, by the ansatz in Eq.~(\ref{eq0m}) if the fields $R_{0}$ and ${\widehat R}_{\hat\mu}$ are
chosen to be linear in $K$.
Even though $K$ must be constant for the EOM to be solved, it should be clear that this
does not imply any constraint on the allowed form of the collective coordinate matrix
$U(t)$ in eq.~(\ref{eq0m}), which can have an arbitrary dependence on time. What we actually
want to do here is to find an appropriate functional dependence of the fields on $U(t)$
such that the time-dependent EOM would be solved if and only if the rotational velocity
$K=-i U^\dagger dU/dt$ was constant.

In order to solve the time-dependent equations (\ref{eomt}) we will consider a 2D ansatz obtained by a
generalization of the cylindrical symmetry of the static case. The ansatz for $R_{\hat\mu}$ and
${\widehat R}_{0}$ is specified by eq.~(\ref{eq0m}) in which the static fields are given by eq.~(\ref{sts}).
Due to the cylindrical symmetry of the static solution the fields in eq.~(\ref{eq0m}) are invariant under 3D space
rotations $x_a\sigma^a\,\rightarrow\,r^\dagger x_a\sigma^ar$ combined with vector $SU(2)$ global transformations
$L,R\rightarrow r\,(L,R)\,r^\dagger$ if $U$ also transforms as $U\rightarrow r^\dagger U r$. We are therefore led
to consider a generalized cylindrical symmetry under which $k_a$ also rotates as the space coordinates do.
Compatibly with this symmetry and with the fact that $R_{0}$ and ${\widehat R}_{\hat\mu}$ must be linear in $K$
we write the ansatz as
\be
\displaystyle
R_{0}({\bf x},z;U)\,=\,U\,{\ov R}_{0}({\bf x},z;K)\,U^\dagger\,+\,i\,U\partial_0U^\dagger \, ,
\;\;\;\;\;\;\;\;\;
{\widehat{R}_{\hat\mu}}({\bf x},z;U)\,=\,{{{\widehat{\ov{R}}}}}_{\hat\mu}({\bf x},z;K)\,,
\label{ansk}
\ee
where
\be
\left\{
\begin{array}{l}
{\ov R}_{0}^a({\bf x},z;K) =\displaystyle \chi_{(x)}(r,z)k_b\Delta^{(x),ab} + v(r,z) (k\cdot\x)\x^a\,\\
\displaystyle
\alpha{{{\widehat{\ov{R}}}}}_{i}({\bf x},z;K) = \displaystyle \frac{\rho(r,z)}{r}\left(k^i - (k\cdot\x)\x^i\right) + B_1(r,z)(k\cdot\x)\x^i + Q(r,z)\epsilon^{ibc}k_b\x_c \,\\
\displaystyle
\rule{0pt}{1.5em}\alpha{{{\widehat{\ov{R}}}}}_{5}({\bf x},z;K) = \displaystyle B_2(r,z)(k\cdot\x) \,
\end{array}
\right.\,.
\label{ansk1}
\ee

It should be noted that the term $i\,U\partial_0U^\dagger=UKU^\dagger$ in eq.~(\ref{ansk}) is purely conventional
as it could have been reabsorbed in the definition of ${\ov R}_{0}$. Nevertheless this choice makes manifest that
our ansatz (\ref{eq0m},\ref{ansk}) can be obtained from the ``barred'' fields in eq.~(\ref{sts},\ref{ansk1}),
which only depend on $U$ through $K$, by performing a time-dependent $SU(2)$ vector gauge transformation with
parameter $U(t)$. This is useful because the action, including the CS term, is invariant under this transformation.
We can therefore obtain the 2D EOM for our ansatz fields by plugging the barred fields, instead of the original ones,
into the 5D EOM. It is important to stress that the ansatz with barred fields is not
truly gauge equivalent to the original one because the
transformation $U(t)$ does not reduce to the identity at the UV boundary, implying that the UV condition
(\ref{uvboundary condition}) is not invariant. Our true ansatz is therefore
provided by eq.~(\ref{eq0m},\ref{ansk}) and
the use of the barred field as we will do in the following is just a useful trick.

At this point it is straightforward to find the zero-mode solution. The EOM for the 2D fields can be obtained
by plugging the ansatz in eq.~(\ref{eomt}), while the conditions at the IR and UV boundaries are derived
from eq.~(\ref{irboundary condition}) and (\ref{uvboundary condition}), respectively. The boundary conditions at $r=0$
are obtained by imposing the regularity of the ansatz, while those for $r\rightarrow\infty$
come from requiring the energy
of the solution to be finite and the topological charge $B$ in eq.~(\ref{Bch}) to be equal to $1$.
More details are presented in appendix~A, where the 2D EOM and the boundary conditions are derived.
Once the 2D equations have been found, however, it is not yet trivial to solve them numerically, the procedure we
followed is described in appendix~B. The reader not interested in detail can however simply accept that a solution
of eq.~(\ref{eomt}) exists and is given by our ansatz for some particular
functional form of the 2D fields which we are able to determine numerically. In
the rest of the paper the 2D fields will always denote this numerical solution of the 2D
equations.

\subsection{The Lagrangian of Collective Coordinates}

The collective coordinate matrix $U(t)$ will be associated with static baryons.
The classical dynamics of the collective coordinates is obtained by plugging eq.~(\ref{eq0m},\ref{ansk}) in the 5D action. One finds $S[U]=\int dt L$ where
\be
L = -M +\frac{\lambda}2\,k_ak^a\,.
\label{eq:Lag}
\ee
The mass $M$ and the moment of inertia $\lambda$ are given respectively by
\be
\begin{array}{l}
\displaystyle
M=8\pi M_5\int_0^\infty dr\int^{z_{\rm IR}}_{z_{\rm UV}}dz\,\left\{a(z)\left[
|D_{\bar\mu}\phi|^2+\frac{1}{4}r^2 A^{2}_{\bar\mu\bar\nu}
+\frac{1}{2r^2}\left(1-|\phi|^2\right)^2-\frac12\left(\partial_{\bar\mu} s\right)^2\right]\right.\\
\displaystyle\left.
-\frac{\gamma L}2
\frac{s}{r}\epsilon^{\bar\mu\bar\nu}
\bigg[\partial_{\bar\mu}(-i \phi^*D_{\bar\nu}\phi+h.c.)
+A_{\bar\mu\bar\nu}\bigg]
\right\}
\,,
\end{array}
\label{eq:mass}
\ee
and
\be
\begin{array}{l}
\displaystyle
\lambda\,=\,16\pi M_5\frac13\int_0^\infty dr\int^{z_{\rm IR}}_{z_{\rm UV}}dz\,\left\{a(z)\left[
-\left(D_{\bar\mu}\rho\right)^2
-r^2\left(\partial_{\bar\mu} Q\right)^2
-2Q^2
-\frac{r^2}4B_{{\bar\mu}{\bar\nu}}B_{{\bar\mu}{\bar\nu}}\right.\right.\\
\displaystyle
\left.\left.
+r^2\left(D_{\bar\mu}\chi\right)^2+
\frac{r^2}2\left(\partial_{\bar\mu}v\right)^2
+\left(\chi_{(x)}\chi_{(x)}+v^2\right)\left(1+\phi_{(x)}\phi_{(x)}\right)
-4v\phi_{(x)}\chi_{(x)}
\right]\right.\\
\displaystyle
\left.
+\gamma L
\bigg[
-2\epsilon^{\bar\mu\bar\nu}D_{\bar\mu}\rho\,\chi_{(x)}\left(D_{\bar\nu}\phi\right)_{(x)}
+2\epsilon^{\bar\mu\bar\nu}\partial_{\bar\mu}\left(r\,Q\right)\,\chi_{(x)}\epsilon^{(xy)}\left(D_{\bar\nu}\phi\right)_{(y)}\right.\\
\displaystyle
\left.
-v\left(\frac12\epsilon^{\bar\mu\bar\nu}B_{\bar\mu\bar\nu}\left(\phi_{(x)}\phi_{(x)}-1\right)
+r\,Q\epsilon^{\bar\mu\bar\nu}A_{\bar\mu\bar\nu}
\right)
+\frac{2r\,Q}{\alpha^2}\epsilon^{\bar\mu\bar\nu} D_{\bar\mu}\rho\partial_{\bar\nu}\left(\frac{s}{r}\right)
\bigg]
\right\}\,.
\end{array}
\label{eq:lambda}
\ee
The notations used in the equations above  are defined in appendix~A;
the covariant derivative symbols, in particular, are associated with two Abelian
residual gauge symmetries which our 2D ansatz has.
Here we simply want to show that $M$ and $\lambda$ could be easily computed, at a given point of the
parameter space, once the numerical solution for the 2D fields is known, by performing a numerical 2D integral.

Let us give some more detail on this theory. For now we proceed at the classical level and we will discuss
the quantization in the next section. Our Lagrangian can be rewritten as
\be
L\,=\, -M + \lambda {\rm Tr}\left[{\dot{U}}^\dagger {\dot{U}}\right]\,=\,-M\,+\,2\lambda \sum_{i}{\dot u}_{i}^2\,,
\label{ccL}
\ee
where we have parametrized the collective coordinates matrix $U$ as $U=u_0\I+i\,u_i\sigma^i$, with
$\sum_{i}{u_i}^2=1$. The Lagrangian (\ref{ccL}) is the one of the classical spherical rigid rotor. The variables
$\{u_0,u_i\}$ are restricted to the unitary sphere $S^3$, which is conveniently parametrized by the coordinates
$q^\alpha\equiv\{x,\phi_1,\phi_2\}$ --which run in the $x\in [-1,1]$, $\phi_1\in [0,2\pi)$ and $\phi_2\in [0,2\pi)$
domains-- as
\be
\displaystyle
u_1+i\,u_2\,\equiv\,z_1\,=\,\sqrt{\frac{1-x}{2}}e^{i\,\phi_1}\,,
\;\;\;\;\;
u_0+i\,u_3\,\equiv\,z_2\,=\,\sqrt{\frac{1+x}{2}}e^{i\,\phi_2}\,,
\label{coord}
\ee
where we also introduced the two complex coordinates $z_{1,2}$. We can now rewrite the Lagrangian as
\be
L\,=\,-M\,+\,2\lambda\,g_{\alpha\beta} {\dot q}^\alpha{\dot q}^\beta\,,
\label{lag}
\ee
where $g$ is the metric of $S^3$ which reads in our coordinates
\be
ds^2\,=\,g_{\alpha\beta}dq^\alpha dq^\beta\,=\,\frac14\frac1{1-x^2}\,dx^2+\frac{1-x}2\,d{\phi_1}^2+\frac{1+x}2\,d{\phi_2}^2\,.
\ee
The conjugate momenta are $p_{\alpha}=\partial L/\partial {\dot q}^\alpha=4\lambda g_{\alpha\beta}{\dot q}^\beta$
and therefore the classical Hamiltonian is
\be
H_c\,=\,M\,+\,\frac1{8\lambda} p_{\alpha}g^{\alpha\beta}(q)p_{\beta}\,.
\label{cham}
\ee

It should be noted that the points $U$ and $-U$ in what we denoted as the space of collective
coordinates actually describe the same field configuration (see eq.~(\ref{eq0m},\ref{ansk})).
The $SU(2)=S_3$ manifold we are considering is actually the universal covering of the collective
coordinate space which is given by $S_3/Z_2$. This will be relevant when we will discuss the
quantization.

\section{Static properties of Nucleons}

\subsection{Skyrmions Quantization}

We should now quantize the classical theory described above, by replacing as usual the classical momenta
$p_\alpha$ with the differential operator $-i\partial/\partial q^\alpha$ acting on the wave functions $f(q)$.
Given that the metric depends on $q$, however, there is an ambiguity in how to extract the quantum  Hamiltonian
$H_q$ from the classical one in eq.~(\ref{cham}). This ambiguity is resolved by requiring the quantum theory to
have the same symmetries that the classical one had. At the classical level, we have an $SO(4)\simeq
SU(2)\times SU(2)$
symmetry under $U\rightarrow U\cdot r^\dagger$ and $U\rightarrow g\cdot U$ with $r,g\in SU(2)$. These correspond,
respectively, to rotations in space and to isospin ({\it i.e.} global vector) transformations, as one can see from
the ansatz in eq.s~(\ref{eq0m},\ref{ansk}).
This is because $K$ is invariant under left multiplication by $g$,
and that the ansatz is left unchanged by performing a rotation
$x_a\sigma^a\rightarrow r^\dagger x_a\sigma^a r$ and
simultaneously sending $U\rightarrow U\cdot r$.
The spin and isospin operators must be given, in the quantum theory, by the generators of these transformations
on the space of wave functions $f(q)$ which are defined by
\be
\left[S^a,U\right]=U\sigma^a/(2)\,,\;\;\;\;\;\left[I^a,U\right]=-\sigma^a/(2)U\,.
\label{comm}
\ee
After a straightforward calculation one finds
\be
\begin{array}{l}
\left\{
\begin{array}{l}
\displaystyle
S^3\,=\,-\frac{i}2\left(\partial_{\phi_1}+\partial_{\phi_2}\right)\\
\displaystyle
S^+\,=\,\frac1{\sqrt{2}}e^{i(\phi_1+\phi_2)}\left[i\sqrt{1-x^2}\partial_x+\frac12\sqrt{\frac{1+x}{1-x}}\partial_{\phi_1}-\frac12\sqrt{\frac{1-x}{1+x}}\partial_{\phi_2}\right]\\
\displaystyle
S^-\,=\,\frac1{\sqrt{2}}e^{-i(\phi_1+\phi_2)}\left[i\sqrt{1-x^2}\partial_x-\frac12\sqrt{\frac{1+x}{1-x}}\partial_{\phi_1}+\frac12\sqrt{\frac{1-x}{1+x}}\partial_{\phi_2}\right]
\end{array}
\right.\\
\left\{
\begin{array}{l}
\displaystyle
I^3\,=\,-\frac{i}2\left(\partial_{\phi_1}-\partial_{\phi_2}\right)\\
\displaystyle
I^+\,=\,-\frac1{\sqrt{2}}e^{i(\phi_1-\phi_2)}\left[i\sqrt{1-x^2}\partial_x+\frac12\sqrt{\frac{1+x}{1-x}}\partial_{\phi_1}+\frac12\sqrt{\frac{1-x}{1+x}}\partial_{\phi_2}\right]\\
\displaystyle
I^-\,=\,-\frac1{\sqrt{2}}e^{-i(\phi_1-\phi_2)}\left[i\sqrt{1-x^2}\partial_x-\frac12\sqrt{\frac{1+x}{1-x}}\partial_{\phi_1}-\frac12\sqrt{\frac{1-x}{1+x}}\partial_{\phi_2}\right]
\end{array}
\right.
\end{array}
\label{spiso}
\ee
where the raising/lowering combinations are $S^{\pm}=(S^1\pm iS^2)/\sqrt{2}$.

The operators in eq.~(\ref{spiso}) should obey the Hermiticity conditions
$\left(S^3\right)^{\dagger}=S^3$, $\left(S^+\right)^{\dagger}=S^-$, and analogously for the isospin.
In order for the Hermiticity conditions to hold we choose the scalar product to be
\be
\langle A|B\rangle\,\equiv\,\int d^3q\,\sqrt{g}{f_A}^{\dagger}(q)f_B(q)\,,
\label{SC}
\ee
where $\sqrt{g}=1/4$ in our parametrization of $S_3$. The reason why this choice of the scalar product gives the
correct Hermiticity conditions is that $S^a$ and $I^a$ (where $a=1,2,3$) can be written as $X^\alpha\partial_\alpha$
with $X^\alpha$ Killing vectors of the appropriate $S_3$ isometries.
The Killing equation $\nabla_\alpha X_\beta+\nabla_\beta X_\alpha=0$ ensures
the generators to be Hermitian with respect to the scalar product (\ref{SC}).

Knowing that the scalar product must be given by eq.~(\ref{SC}) greatly helps in guessing what the quantum
Hamiltonian, which has to be Hermitian, should be. We can multiply and divide by $\sqrt{g}$ the kinetic term
of $H_c$ and move one $\sqrt{g}$ factor to the left of $p_\alpha$. Then we apply the quantization rules and
find
\footnote{The last equality holds because $H_q$ is supposed to be acting on the wave functions,
which are scalar functions.}
\be
H_q\,=\,M-\frac1{8\lambda}\frac1{\sqrt{g}}\partial_\alpha \left(\sqrt{g}g^{\alpha\beta}\partial_\beta\right)\,=\,M-\frac1{8\lambda}\nabla_\alpha\nabla^\alpha\,,
\label{qham}
\ee
which is clearly Hermitian. We can immediately show that $H_q$ commutes with spin and isospin, so that the
quantum theory is really symmetric as required: a straightforward calculation gives indeed
\be
H_q\,=\,M+\frac1{2\lambda}S^2\,=\,M+\frac1{2\lambda}I^2\,.
\ee

It would not be difficult to solve the eigenvalue problem for the Hamiltonian (\ref{qham}), but in order to find
the nucleon wave functions it is enough to note that the versor of $n$-dimensional Euclidean space provides the
$n$ representation of the $SO(n)$ isometry group. In our case, $n=4=(2,2)$, which is exactly the
spin/isospin
representation in which nucleons live. It is immediately seen that $z_1$, as defined in eq.~(\ref{coord}),
has $S^3=I^3=1/2$. Acting with the lowering operators we easily find the wave functions
\be
\displaystyle
\begin{array}{ll}
 |p\,\uparrow\rangle=\displaystyle\frac1{\pi}z_1\,,\;\;\;\;\; & |n\,\uparrow\rangle=\displaystyle\frac{i}{\pi}z_2\,,\\
 |p\,\downarrow\rangle=\displaystyle-\frac{i}{\pi}{\ov z}_2\,,\;\;\;\;\;& |n\,\downarrow\rangle=\displaystyle-\frac{1}{\pi}{\ov z}_1\,,
\end{array}
\ee
which are of course normalized with the scalar product (\ref{SC}). The mass of the nucleons is
therefore $E=M+3/(8\lambda)$.

Notice that the nucleon wave functions are odd under $U\rightarrow -U$, meaning that they are
double-valued on the genuine collective coordinate space $S_3/Z_2$. This corresponds, following
\cite{Finkelstein:1968hy}, to quantize the skyrmion as a fermion and explains how we could get
spin-$1/2$ states after a seemingly bosonic quantization without violating spin--statistic.

Let us now summarize some useful identities which will be used in our calculation. First of all, it is
not hard to check that, after the quantization is performed the rotational velocity becomes
\be
k^a\,=\,-i\,{\rm Tr}\left[U^{\dagger}{\dot U}\sigma^a\right]\,=\,\frac1\lambda S^a\,,
\ee
and analogously
\be
i\,{\rm Tr}\left[{\dot U}U^{\dagger}\sigma^a\right]\,=\,\frac1\lambda I^a\,.
\label{Iden}
\ee
Other identities which we will use in our calculations are
\bea
\displaystyle
&&\langle {\rm Tr}\left[U\,\sigma^b U^\dagger\sigma^a\right] = -\frac83 S^b I^a\rangle\,, \nn\\
&&\langle {\rm Tr}\left[U\,\sigma^b \x_b (k\cdot\x) U^\dagger\sigma^a\right] = -\frac2{3\lambda} I^a\rangle\,,
\label{qr}
\eea
where the VEV symbols $\langle...\rangle$ mean that those are not operatorial identities, but they only hold when
the operators act on the subspace of nucleon states. Notice that the second equation in (\ref{qr}) is implied by
the first one if one also uses the commutation relation (\ref{comm}), eq.~(\ref{Iden}) and the fact that, on
nucleon states, $\langle\left\{S^a,S^i\right\}=\delta^{ai}/2\rangle$.

\subsection{The Nucleon Form Factors}

The nucleon form factors parametrize the matrix element of the currents on two
nucleon states. For the isoscalar and isovector currents we have
\bea\label{eqEMCurrCorr}
\displaystyle
\langle N_f(p') | J^\mu_{S}(0) | N_i(p)\rangle = \bar u_f(p') \left[
F_1^S(q^2) \gamma^\mu + \frac{i F_2^S(q^2)}{2 M_N} \sigma^{\mu\nu}q_\nu\right]u_i(p),\nn\\
\langle N_f(p') | J^{\mu a}_{V}(0) | N_i(p)\rangle = \bar u_f(p') \left[
F_1^V(q^2) \gamma^\mu + \frac{i F_2^V(q^2)}{2 M_N} \sigma^{\mu\nu}q_\nu\right]\left(2I^a\right) u_i(p),
\eea
where the currents are
defined as $J_{V}^a=J_R^a+J_L^a$ and $J_S=1/3\left({\widehat J}_R+{\widehat J}_L\right)$
in terms of the chiral ones. In the equation above
$q \equiv p'-p$ is the $4$-momentum transfer, $N_i$ and $N_f$ are the initial and final
nucleon states and $u_i(p)$, $\bar u_f(p')$ their wave functions,
$I^a=\sigma^a/2$ is the isospin generators and
$\sigma^{\mu\nu} \equiv i/2 [\gamma^\mu, \gamma^\nu]$.
For the axial current $J_{A}^a=J_R^a-J_L^a$ we have
\be\label{eqAxCurrCorr}
\displaystyle
\langle N_f(p') | J^a_{A \mu}(0) | N_i(p)\rangle = \bar u_f(p')
G_A(q^2) \left[ \gamma_\mu -\frac{2 M_N}{q^2} q^\mu \right]\gamma^5 I^a u_f(p)\,.
\ee
Exact axial and isospin symmetries, which hold in our model, have been assumed in the
definitions above.

In our non-relativistic model the current correlators will be computed in the Breit frame
in which the initial nucleon
has 3-momentum $-{\vec q}/2$ and the final $+{\vec q}/2$ (i.e. $p^\mu = (E, -{\vec q}/2)$ and
$p'^\mu = (E, {\vec q}/2)$, and $q^2=-{\vec q\,}^2$, with $E=\sqrt{M_N^2 + {\vec q\,}^2/4}$).
Notice that the textbook definitions in eq.s~(\ref{eqEMCurrCorr},\ref{eqAxCurrCorr})
involve nucleon states which are normalized with $\sqrt{2 E}$; in order to match with
our non-relativistic normalization we have to divide all correlators by $2M_N$. The
vector currents become
\bea
\displaystyle
\langle N_f({\vec q}/2) | J^0_{S}(0) | N_i(-{\vec q}/2)\rangle &=&  G_{E}^{S}({\vec q\,}^2) \chi_f^\dagger \chi_i\,,\nn\\
\displaystyle
\langle N_f({\vec q}/2) | J^i_{S}(0) | N_i(-{\vec q}/2)\rangle &=& i\, \frac{G_M^{S}({\vec q\,}^2)}{2 M_N} \chi_f^\dagger 2 ({\vec S} \times {\vec q})^i \chi_i\,,\nn\\
\displaystyle
\langle N_f({\vec q}/2) | J^{0 a}_{V}(0) | N_i(-{\vec q}/2)\rangle &=& G_{E}^{V}({\vec q\,}^2) \chi_f^\dagger \left(2I^a\right) \chi_i\,,\nn\\
\displaystyle
\langle N_f({\vec q}/2) | J^{i a}_{V}(0) | N_i(-{\vec q}/2)\rangle &=& i\, \frac{G_M^{V}({\vec q\,}^2) }{2 M_N}
\chi_f^\dagger 2 ({\vec S}  \times {\vec q})^i \left(2I^a\right) \chi_i\,,
\label{bf}
\eea
where we defined
\be
G_E^{S,V}(-q^2) = F_1^{S,V}(q^2) + \frac{q^2}{4 M_N^2} F_2^{S,V}(q^2)\,, \qquad G_M^{S,V}(-q^2) = F_1^{S,V}(q^2) + F_2^{S,V}(q^2)\,,
\ee
and used the definition
$({\vec S} \times {\vec q})^i \equiv \varepsilon^{ijk} S^j q^k$.
The nucleon spin/isospin vectors of state $\chi_{i,f}$  are
normalized to $\chi^\dagger \chi = 1$.
For the axial current we find
\bea
\langle N_f({\vec q}/2) | J_A^{i,a}(0) | N_i(-{\vec q\,}/2)\rangle &=&
\chi_f^\dagger \frac{E}{M_N} G_A({\vec q\,}^2) 2 S^i_T
I^a\chi_i\,,\nn\\
\langle N_f({\vec q\,}/2) | J_A^{0,a}(0) | N_i(-{\vec q\,}/2)\rangle &=&0
\label{bfa}
\eea
where ${\vec S}_T \equiv {\vec S} - \hat{{\vec{q}}}\  {\vec S}\cdot{\hat{\vec{q}}}$ is the transverse  component of the spin operator.

It is straightforward to compute the matrix elements of the currents in position space on
static nucleon states. Plugging the ansatz (\ref{sts},\ref{eq0m},\ref{ansk1},\ref{ansk}) in
the definition of the currents (\ref{cur0}) and performing the quantization one obtains
quantum mechanical operators acting on the nucleons. The matrix elements are easily
computed using the results of sect.~3.1. We finally obtain the form factors by taking the
Fourier transform and comparing with eq.s~(\ref{bf},\ref{bfa}). We have
\footnote{It is quite intuitive that the form factors can be computed in this way.
Given that solitons are infinitely heavy at small coupling, in the Breit frame
they are almost static during the process of scattering with the current.
To check this, however, we should perform the
quantization of the collective coordinates associated with the center-of-mass motion, as it
was done in \cite{Braaten:1986iw} for the original 4D Skyrme model.}
\bea
\displaystyle
&&G_E^S\,=\,-\frac{N_c}{6\pi\gamma L}\int dr\,r\,j_0(qr)\left(a(z)\partial_zs\right)_{UV}\nn\\
&&G_E^V\,=\,\frac{4\pi M_5}{3\lambda}\int dr\,r^2\,j_0(qr)\left[a(z)\left(\partial_zv-2\left(D_z\chi\right)_{(2)}\right)\right]_{UV}\nn\\
&&G_M^S\,=\,\frac{8\pi M_NM_5\alpha}{3\lambda}\int dr\,r^3\,\frac{j_1(qr)}{qr}\left(a(z)\partial_zQ\right)_{UV}\nn\\
&&G_M^V\,=\,\frac{M_N\,N_c}{3\pi L\gamma \alpha}\int dr\,r^2\,\frac{j_1(qr)}{qr}\left(a(z)\left(D_z\phi\right)_{(2)}\right)_{UV}\nn\\
&&G_A\,=\,\frac{M_N}{E}
\frac{N_c}{3\pi\alpha\gamma L}
\int dr\,r\left[
a(z)\frac{j_1(qr)}{qr}
\left(\left(D_z\phi\right)_{(1)}-r\,A_{zr}\right)
-a(z)\left(D_z\phi\right)_{(1)} j_0(qr)
\right]_{UV}
\label{cff}
\eea
where $j_n$ are spherical Bessel functions which arise because of the Fourier transform.

\section{Results}

In this section we will present our results. After discussing some qualitative features,
such as the large-$N_c$ scaling of the form factors and the divergences of the isovector radii due to exact
chiral symmetry, we extrapolate to the physically relevant case of $N_c=3$ and perform a quantitative
comparison with the experimental data. Consistently with our working hypothesis that the 5D model really
describes large-$N_c$ QCD we find a $30\%$ relative discrepancy.

\subsubsection*{Large-$N_c$ Scaling}

Let us take all the three parameters $\alpha$, $\gamma$ and $L$ of our 5D model to scale like $N_{c}^0$ for
large-$N_c$. Eq.~(\ref{gamma}) therefore implies that the coupling $M_5$ grows like $N_c$ and the semiclassical
expansion in 5D coincides with the $1/N_c$ expansion on the 4D side. Notice that these scaling of the parameters
are uniquely dictated by what we know to be the large-$N_c$ scaling of meson couplings and masses.
In the baryon sector, the solitonic solution is independent of $N_c$ given that $M_5$ factorizes out of the action
and does not appear in the EOM. The classical mass $M$ and the moment of inertia $\lambda$ therefore scale like $N_c$
and the scaling of the form factors can be easily read from eq.~(\ref{cff}).

In large-$N_c$ QCD the scaling of several baryon observables is known \cite{Witten:1979kh}. The mass grows with $N_c$
as in our model, but this is expected to be a common feature of any soliton model. The matrix element of currents on
normalized nucleon states should be of the form $N_{c}^pF(q^2)$ with $p=1$ even though cancellations, {\it i.e.}
$p<1$, are not excluded. All the radii should therefore scale like $N_{c}^0$ and this is what we find in our model.
We also find the ``naive" --{\it{i.e.}} with $p=1$-- overall scaling for the electric scalar ($G_{E}^S$), magnetic
vector ($G_{M}^V$) and axial ($G_A$) form factor; notice that, due to the definition in eq.~(\ref{bf}), the magnetic
form factors scale with one more power of $N_c$ than what the corresponding current matrix element does. We however
find two cancellations: due to the $1/\lambda$ factor the electric vector $G_{E}^V$ and the magnetic scalar $G_{M}^S$
scale like $N_{c}^0$ and $N_{c}^1$, respectively. This corresponds to
a ``$p=0$'' scaling of the associated currents.

The reason for the cancellation in $G_{E}^V$ is very simple to understand. Remembering that the temporal component
of the current at zero momentum gives the conserved charge and looking at the definitions (\ref{bf}) one immediately
obtains two consistency conditions: $G_{E}^S(0)=N_c/6$, because in the nucleons there are $N_c$ quarks which have
$U(1)_V$ charge $1/6$ each in our conventions, and $G_{E}^V(0)=1/2$, because nucleons are in the $1/2$ representation
of isospin. It is not difficult to see that these consistency conditions are respected by our model as they are implied
by the EOM, and they are fulfilled to great accuracy ($0.1\%$) by the numerical solution. The above discussion
implies, in particular, that while the electric scalar form factor $G_{E}^S$ has the naive $N_c$ scaling, the electric
vector $G_{E}^V$ does not.

We are not able to prove that the cancellation in $G_{M}^S$ actually takes place in large-$N_c$ QCD, but we can
check that it occurs in the naive quark model, or better in its generalization for arbitrary odd $N_c=2\,k+1$
\cite{Witten:1983tx}. In this non-relativistic model the Nucleon wave function is made of $2k+1$ quark states, $2k$
of which are collected into $k$ bilinear spin/isospin singlets while the last one has free indices which give to the
Nucleon its spin/isospin quantum numbers. Of course, the wave function
is symmetrized in flavor and spin given that the color indices are contracted with the antisymmetric tensor and the
spatial wave function is assumed to be symmetric. The current operator is the sum of the currents for the $2k+1$
quarks, each of which will assume by symmetry the same form as in eq.~(\ref{bf}). If $S_{1,2}$ and $I_{1,2}$
represent the spin and isospin operators on the quarks $q_{1,2}$ the
operators $S_1+S_2$ and $I_1+I_2$ will vanish on the
singlet combination of the two quarks, but $S_1I_1+S_2I_2$ will not.
The $k$ singlets will therefore only contribute to $G_{E}^S$, $G_{M}^V$ and $G_{A}$, which will have the
naive scaling, while for the others we find cancellations.

A detailed calculation can be found in \cite{Karl:1984cz} where, among other things, the proton and neutron
magnetic moments and the axial coupling are computed in the naive quark model. The magnetic moments are related
to the form factor at zero momentum as $\mu_V/\mu_N=G_{M}^V(0)$ and $\mu_S/\mu_N=G_{M}^S(0)$ where $\mu_N=1/(2M_N)$
is the nuclear magneton and $2\,\mu_V=\mu_p-\mu_n$, $2\,\mu_S=\mu_p+\mu_n$. In accordance with the previous discussion,
the results in the naive quark model are $2\mu_S=\mu_u+\mu_d$ and $2\mu_V=2k/3(\mu_u-\mu_d)$, where $\mu_{u,d}$
are the quark magnetic moments, while for the axial coupling one finds 
$g_A=G_A(0)=2k/3+1=N_c/3+2/3$ which scales like $N_c$
as expected. Notice that the
$1/N_c$ corrections to the axial coupling are quite big in
the naive quark model: for $N_c=3$ the leading term contributes as $1$ while the ``true'' results
is $5/3$, which is $67\%$ bigger. 
We have of course no reason to believe that such big corrections should
persist in the true large-$N_c$ expansion of QCD, this trivial remark simply suggests that
``large'' $1/N_c$ corrections to the form factors are not excluded.

\subsubsection*{Divergences in the Chiral Limit}

It is well known that in QCD the isovector electric $\langle r_{E,\,V}^2\rangle$ and magnetic
$\langle r_{M,\,V}^2\rangle$ radii which are proportional, respectively, to the $q^2$ derivative of $G_{E}^V$
and $G_{M}^V$ at zero momentum, diverge in the chiral limit \cite{Beg:1973sc}. If the nucleons are effectively
described by an isospin doublet of point-like spinors added to the $\chi_{\bf PT}$ Lagrangian this effect
comes from pion loops which are IR divergent in the massless pion limit $m_\pi\rightarrow0$ \cite{Gasser:1987rb}.
It is not obvious, however, that the divergences should survive in the large-$N_c$ limit, {\it{i.e.}} that they
should already appear in the leading term of the perturbative $1/N_c$ expansion. This is so because describing
baryons as weakly coupled particles, which is a reasonable approximation in real-world $N_c=3$ QCD, is not possible
at large-$N_c$ given that their coupling with pions grows like $N_{c}^{3/2}$. It was noticed in
\cite{Dashen:1993as} that, in a model which only contains the nucleon doublet, the
pion--nucleon scattering amplitude grows like $N_c$ violating unitarity and also
contradicting the usual large-$N_c$ counting rules. The theory is therefore inconsistent
and the full infinite tower of large-$N_c$ baryons must be added. Moreover, it is very easy
to see that the results of \cite{Beg:1973sc,Gasser:1987rb} are not compatible with
large $N_c$: the one-loop corrections to the radii, which are of course finite for finite
$m_\pi$, have the wrong scaling and grow like $N_{c}$. Given that we cannot apply the
results of \cite{Beg:1973sc,Gasser:1987rb}, we cannot conclude that the radii must diverge
in our model at the leading order in the semiclassical expansion of the soliton, but there
is of course no problem if they do. We must however check that all the other radii are
finite, and this is what we will do in the following. What we will find is the same as in
the 4D Skyrme model \cite{Adkins:1983ya}: all radii and form factors are finite but the
electric and magnetic isovector ones.

In our model, as in the Skyrme model, divergences in the integrals of eq.~(\ref{cff}) which define the form
factors are due, as in QCD, to the massless pions. If all the fields were massive, indeed, any solution to the
EOM would fall down exponentially at large $r$ while in the present case power-like behaviors can appear. These
power-like terms in the large-$r$ expansion of the solution can be derived analytically by performing a Taylor
expansion of the fields around infinity ($1/r=0$), substituting into the EOM and solving order by order in $1/r$.
The exponentially suppressed part of the solution will never contribute to the expansion. In the gauge in
which (the form factors are, of course, gauge invariant) the topological twist is at the origin $r=0$ and the solution
is trivial for $r\rightarrow \infty$ the first few terms are \footnote{In the equations which follow we put $L=1$ for simplicity.}
\be
\left\{
\begin{array}{l}
\displaystyle A_1 = \frac{2 z (z-1)}{r^3} \beta\\
\displaystyle A_2 = \frac{\beta}{r^2} + \frac{4 z^3 - 6 z^2 +1}{2 r^4} \beta\\
\displaystyle \phi_1 = \frac{z (1-z)}{r^2} \beta + \frac{z(z^3 - 2 z^2 +1)}{2 r^4}\beta\\
\displaystyle \phi_2 = -1 + \frac{z^2 (3-2z)}{2 r^4} \beta^2\\
\displaystyle s = \frac{z^2(z^6 - 4 z^4 +8)}{4 r^8} \gamma \beta^3
\end{array}
\right.
\qquad
\left\{
\begin{array}{l}
\displaystyle \chi_1 = \frac{z (z-1)}{r^2} \beta\\
\displaystyle \chi_2 = 1 + \frac{z^2 (2 z-3)}{2 r^4} \beta^2\\
\displaystyle v = -1 + \frac{z^2 (z^2-3)^2}{12 r^6} \beta^2\\
\displaystyle q = - \frac{z^2 (z^6 - 4 z^4+8)}{4 r^8} \gamma \beta^3
\end{array}
\right.
\label{larger}
\ee
where $\beta$ is an unknown parameter which depends on the entire solution and can only be determined
numerically. We checked that the large-$r$ behavior of our numerical solution is very well
approximated by eq.~(\ref{larger}).
Substituting these expressions into the definitions of the form factors (\ref{cff})
one gets
\be
\left\{\displaystyle
\begin{array}{l}
\displaystyle G_E^S\;\propto\; \displaystyle\beta^3 \int dr \frac{1}{r^7} j_0(q r)+\ldots\\
\displaystyle G_E^V \;\propto\; \displaystyle\beta^2 \int dr \frac{1}{r^2} j_0(qr)+\ldots\\
 G_M^S\;\propto\; \displaystyle\beta^3 \int dr \frac{1}{r^5}  \frac{j_1(qr)}{qr}+\ldots\\
G_M^V\;\propto\; \displaystyle\beta^2 \int dr \frac{1}{r^2}   \frac{j_1(qr)}{qr}+\ldots
\end{array}
\right.\,.
\label{eq:largerbehaviour}
\ee
All the form factors are finite for any $q$, including $q=0$.
The electric and magnetic radii, however, are defined as
\be
\langle r^2_{E,M} \rangle = -\frac{6}{G_{E,M}(\vec q\,^2 = 0)}
\left.\frac{d G_{E,M}(\vec q\,^2)}{d \vec q\,^2}\right|_{\vec q\,^2=0}\,,
\label{eq:radii}
\ee
and taking a $q^2$ derivative of eq.s~(\ref{eq:largerbehaviour}) makes one more power of $r^2$ appear
in the integral. It is easy to see that the scalar radii are finite, while the vector ones are divergent
as anticipated. We will now discuss the axial coupling and the axial radius and show that both are finite.

The expression in eq.~(\ref{cff}) for the axial form factor $G_A$
presents some subtleties for vanishing $q^2$.
Given the asymptotic expansion of the solution in eq.~(\ref{larger}) the
axial coupling integral behaves for large $r$ like
\be
\displaystyle G_A\;\propto\;
\int dr\left[
 \left(\frac{3}{r} \beta - \frac{1}{r^5} \beta^3\right)\frac{j_1(qr)}{qr}
+ \left(-\frac{1}{r} \beta + \frac{5}{7 r^5} \beta^3\right) j_0(qr)+\ldots\right]\,,
\label{axas}
\ee
where the leading $1/r$ terms (the ones which are linear in $\beta$) can be obtained from
eq.~(\ref{larger}) while for the others one needs higher order terms which are not reported
in eq.~(\ref{larger}). The integral in eq.~(\ref{axas}) is finite integral for any $q\neq0$.
For $q \rightarrow 0$, however, the integral
is not uniformly convergent and one cannot exchange the limit with the integration. The leading
$1/r$ term in eq.~(\ref{axas}) is indeed given by
$I(q) = \beta\int_0^\infty dr\ (1/r)\left(3 j_1(qr)/(qr)-j_0(qr)\right)$ which is independent of $q$
and equal to $\beta/3$. Given that the argument of the integral vanishes for $q\rightarrow0$
exchanging the limit and integral operations would give the wrong result $I(0)=0$.
To restore uniform convergence and obtain an analytic formula
for $g_A$ one can subtract the $I(q)$ term from the expression
in eq.~(\ref{cff}) for $G_A$ and replace it with $\beta/3$.
Rewriting the axial form factor in this way is also useful to establish that
the axial radius, which seems divergent if looking at eq.~(\ref{axas}), is
on the contrary finite. The $I(q)$ term, indeed, does not contribute
to the $q^2$ derivative and the ones which are left in eq.~(\ref{axas}) give
a finite contribution.

We have found, in summary, that all the form factors and radii are finite but the isovector ones.
Given that the divergences are related with the large-distance behaviors of the fields, and that our model
reduces to the Skyrme model in the IR, this result is not surprising. A different result has been found, however,
in Ref.~\cite{Hashimoto:2008zw}, where the nucleon form factors have been computed in the Sakai--Sugimoto model.
In that case all the radii are finite. The Sakai--Sugimoto baryons correspond, as explained in the Introduction,
to the small-size limit of the 5D skyrmions we are considering, and we should recover the results of
\cite{Hashimoto:2008zw} if we perform a small-$\gamma$ expansion which correspond to the $1/\lambda$ expansion
considered in \cite{Hashimoto:2008zw}. As $\gamma$ decreases our soliton becomes more and more localized around
$(r=0,z=z_{\IR} )$ and at any large but fixed value of $r$ the deviations from the pure-gauge configuration
become smaller and smaller. The small-$\gamma$ expansion of the asymptotic
solution (\ref{larger}) therefore coincides with the small-$\beta$ expansion. By looking at
eq.~(\ref{eq:largerbehaviour}) we see that the power-like terms in the isoscalar and isovector
form factor densities appear at high orders in $\beta$ and this explains why these densities were
found to be exponentially damped in \cite{Hashimoto:2008zw}. For the axial form factor, as eq.~(\ref{axas})
shows, power-like terms are present at the linear order in $\beta$. The same term has been found in
 \cite{Hashimoto:2008zw} but it does not lead to any divergence as explained in the previous paragraph.

A possible physical explanation of the finiteness of the radii in the Sakai--Sugimoto model is that the 5D
soliton effectively reduces to a 5D particle in the limit of small size, a possibility discussed in
\cite{Hong:2007kx}. For a 5D particle no divergences appear in the radii at the leading order in the semiclassical
expansion ({\it i.e.} at tree level) and the divergences should arise, in analogy with the case of a 4D particle,
at loop level. Following the analogy, however, one could expect the divergent loop corrections (or better the
enhanced loop corrections for small but finite pion mass) to have, as it happens for the 4D particle, the wrong
large-$N_c$ scaling. By the same reasoning one could expect unitarity violation in the pion-nucleon scattering
amplitude at tree-level.

\subsubsection*{Pion Form Factor and Goldberger--Treiman relation}

It is of some interest to define and compute the pion-nucleon form factor which parametrizes the matrix element
on Nucleon states of the pion field. In the Breit frame (for normalized nucleon states)
 it is
\be
\displaystyle
\langle N_f({\vec q}/2) | \pi^{a}(0) | N_i(-{\vec q\,}/2)\rangle =
-\frac{i}{2M_N{\vec q}^2}G_{NN\pi}({\vec q}^2)\chi_f^\dagger(2 S^i) q_i (2 I^a)\chi_i\,,
\label{dme}
\ee
where $\pi^a(x)$ is the normalized and ``canonical" pion field operator. The field is canonical
in the sense that its quadratic effective Lagrangian only contains the canonical kinetic term
${\mathcal L}_2=1/2(\partial\pi_a)^2$, or equivalently that its propagator is the canonical one,
without a non-trivial form factor. With this definition, $G_{NN\pi}$ is the vertex form factor of the
meson-exchange model for nucleon-nucleon interactions \cite{Machleidt:1987hj} and corresponds
to an interaction \footnote{Nucleon scattering, in our model, is a soliton scattering process and we
have no reason to believe that it can be described by meson-exchange, {\it i.e.} that contact terms
are suppressed. Therefore, we will not attempt any comparison of our form factor with the one
used in meson-exchange models.}
\be
{\mathcal L}_{NN\pi}=i\, (G_{NN\pi}(\Box)\pi_a){\ov N}\gamma^\mu\gamma_5(2I^a)N\,.
\ee
On-shell, the form factor reduces to the pion-nucleon coupling constant, $G_{NN\pi}(0)=g_{NN\pi}$,
whose experimental value is $g_{NN\pi}=13.5\pm0.1$.

The pion field which matches the requirements above is given by the zero-mode of the
KK decomposition. In the unitary gauge $\partial_z(a(z) A_5) = 0$,
where $A_M \equiv (L_M-R_M)/2$, and for AdS$_5$ space $a(z)=L/z$ one has
\be
\displaystyle A^{(un)}_5(x, z) = \frac{1}{F_\pi L}\frac1{a(z)} \pi^a(x)\sigma_a\,,
\ee
with $F_\pi^2=2M_5/\int dz/a(z)=4M_5/L$. \footnote{We take the opportunity here to remark that the formula
for $F_\pi$ reported in \cite{Pomarol:2008aa}, though written for general warp factor $a(z)$,
is only correct in the case of AdS$_5$ space in which $a(z)=L/z$.}
Gauge-transforming back to the gauge in which our numerical solution is provided and using
the ansatz in eq.s~(\ref{sts},\ref{eq0m}) we find the pion field
\be
\displaystyle
\pi^a\,=\,-\frac{F_\pi}{2} \int_{z_{\UV}}^{z_{\IR}} dz A_2(r,z){\widehat x}^b
{\rm Tr}\left[U\sigma_bU^\dagger \sigma^a\right]\,.
\ee
Taking the matrix element of the above expression and comparing with eq.~(\ref{dme})
one obtains
\be
G_{NN\pi} (q^2) =  -\frac{8 \pi}{3} M_N F_\pi q
\int_{0}^\infty dr j_1(q r) \int d z\, r^2\,A_2(r,z)\,.
\label{rpc}
\ee

Using eq.~(\ref{larger}) it is easy to understand that the $q\rightarrow0$ limit
of $G_{NN\pi}$ is completely determined by the large-$r$ behavior of the field
$A_2$, and in particular by the leading $\beta/r^2$ term. Due to the $q$ factor, indeed,
only the divergent part of the integral contributes. We then find
\be
g_{NN\pi} = -\frac{2 N_C}{3 \pi} \frac{M_N}{F_\pi \gamma \alpha} \beta\,.
\ee
We used the formula above to check numerically that the Goldberger--Treiman relation
$F_\pi g_{\pi NN}=M_N g_A$ holds in our model, we find that it is verified to $0.01\%$
on our numerical solution.
We can also demostrate the Goldberger--Treiman relation by using eq.~(33,34) of
\cite{Adkins:1983ya} which show that also $g_A$ is determined by the asymptotic
behavior of the axial current. We indeed obtain
\be
g_A = -\frac{2 N_C}{3 \pi \alpha \gamma} \beta\,.
\ee

\subsubsection*{Comparison with Experiments}

Let us now compare our results with real-world QCD, we therefore fix
the number of colors $N_c=3$ and choose our microscopic parameters
to be $1/L \simeq 343\ \text{MeV}$,
$M_5 L \simeq 0.0165$ and $\alpha \simeq 0.94$ ($\gamma \simeq 1.23$).
These values are obtained by minimizing the root mean square error (RMSE)
in the mesonic sector. The detailed list of the observables we used
can be found in \cite{Pomarol:2008aa} and the minimum RMSE for mesons
is found to be $11\%$.

\begin{table}
\centering{
\begin{tabular}{c@{\hspace{2em}}c@{\hspace{2em}}c@{\hspace{2em}}c}
\hline
& Experiment & AdS$_5$ & Deviation\\
\hline
$M_N$ & $940\ \text{MeV}$ & $1130\ \text{MeV}$ & $20\%$\\
$\mu_S$ & $0.44$ & $0.34$ & $30\%$\\
$\mu_V$ & $2.35$ & $1.79$ & $31\%$\\
$g_A$ & $1.25$ & $0.70$ & $79\%$\\
$\sqrt{\langle r_{E,S}^2\rangle}$ & $0.79\ \text{fm}$ & $0.88\ \text{fm}$ & $11\%$\\
$\sqrt{\langle r_{E,V}^2\rangle}$ & $0.93\ \text{fm}$ & $\infty$ \\
$\sqrt{\langle r_{M,S}^2\rangle}$ & $0.82\ \text{fm}$ & $0.92\ \text{fm}$ & $12\%$\\
$\sqrt{\langle r_{M,V}^2\rangle}$ & $0.87\ \text{fm}$ & $\infty$ \\
$\sqrt{\langle r_{A}^2\rangle}$ & $0.68\ \text{fm}$ & $0.76\ \text{fm}$ & $12\%$\\
\hline
$\mu_p/\mu_n$ & $-1.461$ & $-1.459$ & $0.1\%$\\
\hline
\hline
\end{tabular}
\caption{Prediction of the nucleon observables with the microscopic
parameters fixed by a fit on the mesonic observables.
The deviation from the empirical data
is computed using the expression $|th-exp|/\min(|th|, |exp|)$, where
$th$ and $exp$ denote, respectively, the prediction of our model and the
experimental result.}\label{tab:MesonFit}}
\end{table}

The numerical results of our analysis and the deviation with respect to the
experimental data are reported in table~\ref{tab:MesonFit}.
We find a fair agreement with the experiments, a $36\%$ total RMSE
which is compatible with the expected size of $1/N_c$ corrections.
We discussed in the previous section that
the isovector radii are divergent because of the chiral limit,
it would be interesting to add the pion mass to the model and compute these observables.
Table~\ref{tab:MesonFit} also shows the proton-neutron magnetic moment ratio, which is
in perfect agreement with the experimental value. Notice that for this observable,
due to the different scalings of $\mu_S$ and $\mu_V$ with $N_c$, our
computation includes two orders of the $1/N_c$ expansion: the leading order value
which is $-1$ and the next-to-leading $1/N_c$ correction which accounts for the extra
$-0.46$. The axial charge is the one which shows
the larger (almost $100\%$) deviation, and indeed removing this observable the RMSE decreases
to $21\%$. We cannot exclude that, in a theory in which the naive expansion parameter is $1/3$,
enhanced $80\%$ relative corrections to few observables might appear at the next-to-leading order.
This failure in $g_A$, therefore, does not invalidate the general picture.

\begin{figure}[t]
\centerline{ \hspace*{-0.5cm}
\includegraphics[width=0.46 \textwidth]{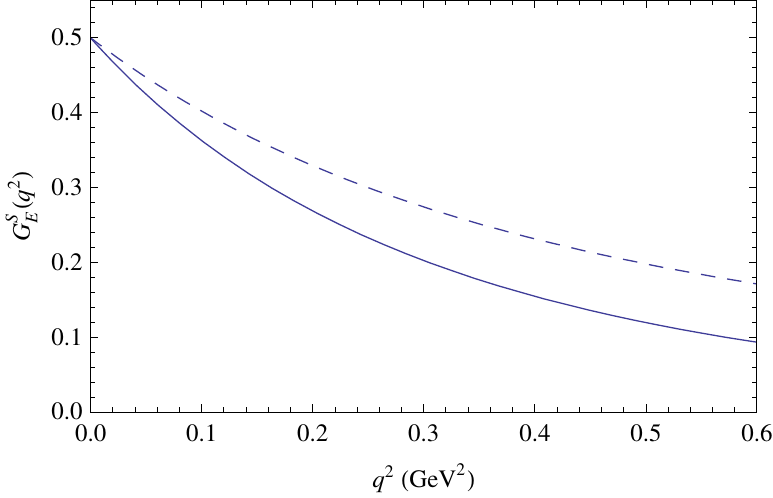}
\hspace*{0.5cm}
\includegraphics[width=0.46 \textwidth]{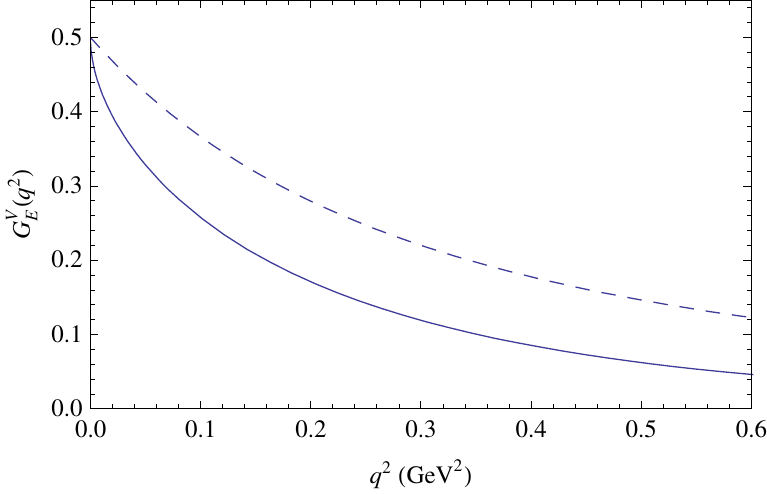}
}
\caption{Scalar (left) and vector (right) electric form factors.
We compare the results with the empirical dipole fit (dashed line) \cite{Meissner:1987ge}.}
\label{Fig:ElectricFormFactors}
\end{figure}

\begin{figure}[t]
\centerline{ \hspace*{-0.5cm}
\includegraphics[width=0.46 \textwidth]{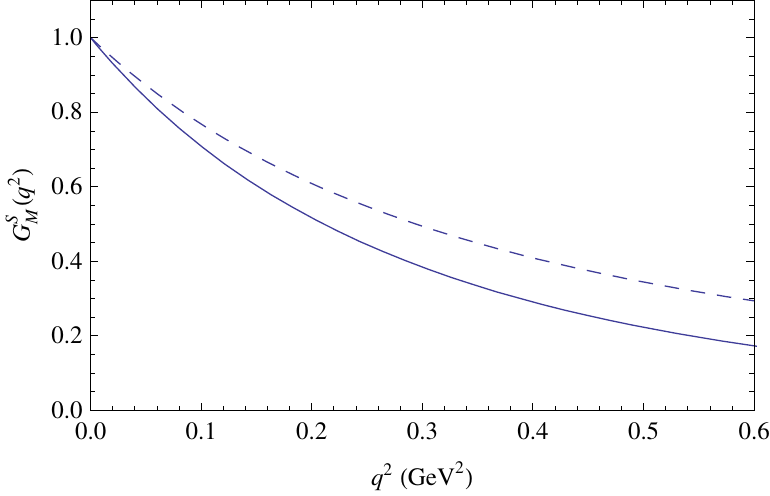}
\hspace*{0.5cm}
\includegraphics[width=0.46 \textwidth]{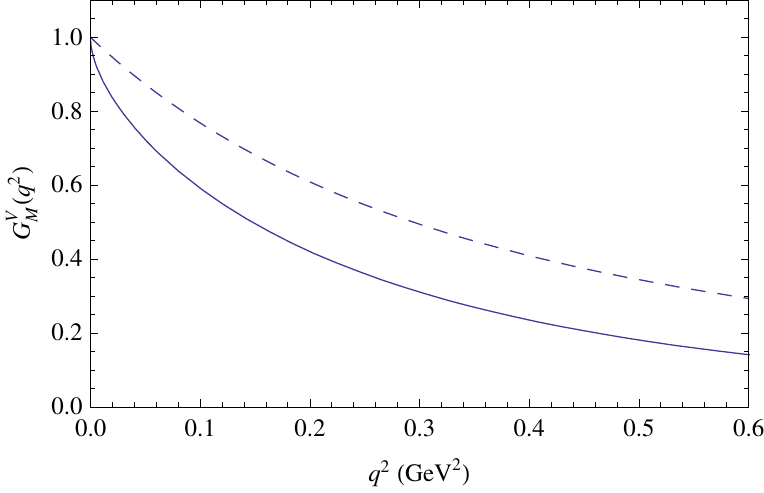}
}
\caption{Normalized scalar (left) and vector (right) magnetic form factors.
We compare the results with the empirical dipole fit (dashed line) \cite{Meissner:1987ge}.}
\label{Fig:MagneticFormFactors}
\end{figure}

\begin{figure}[t]
\centerline{
\includegraphics[width=0.46 \textwidth]{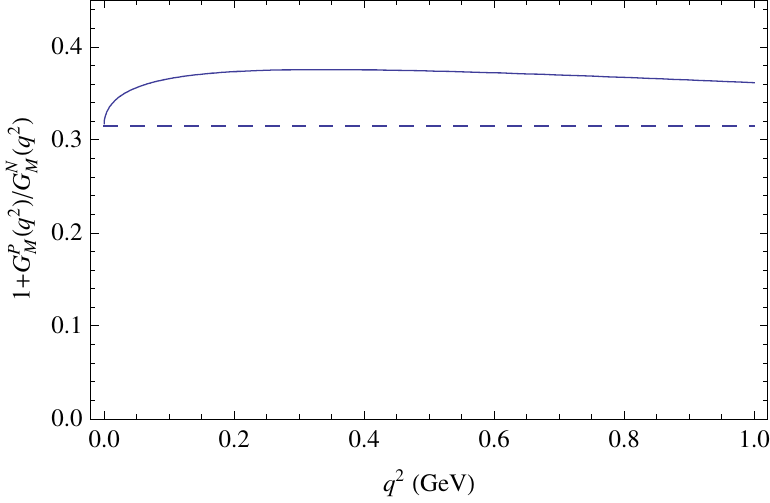}
\hspace*{0.5cm}
\includegraphics[width=0.46 \textwidth]{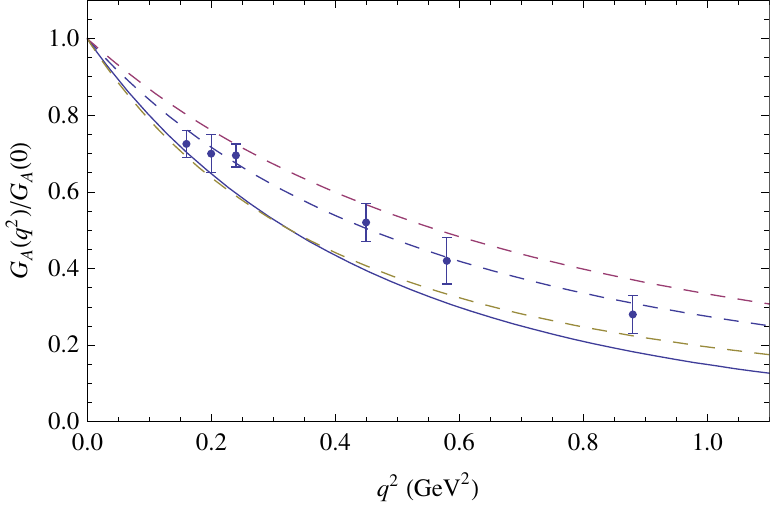}
}
\caption{Left: deviation of the ratio of proton and neutron magnetic form factors
from the large $N_c$ value (solid line), compared with the dipole fit
of the experimental data (dashed line).
Right: normalized axial form factor (solid line) compared with
the empirical dipole fit
(dashed lines) \cite{Meissner:1987ge} and with
the experimental data taken from \cite{Amaldi:1972vf, DelGuerra:1976uj}.}
\label{Fig:AxialFormFactor}
\end{figure}

It is interesting to notice that a much better prediction for $g_A$ is obtained 
if one uses, instead of the standard procedure \cite{Adkins:1983ya} considered in this 
paper, a different approach to the quantization of collective coordinates of the 
skyrmion, which has been proposed 
in Ref.~\cite{Amado:1986ef}. 
The results of Ref.~\cite{Amado:1986ef} can be directly applied to our case since,  
for what concerns the collective coordinate quantization, the 5D nature of our soliton
 is immaterial.
We therefore find that the prediction for $\mu_S$ and for the radii are unaffected while 
both $\mu_V$ and $g_A$ are rescaled by $5/3$. We still obtain a good 
prediction for $\mu_V=2.98$ (which is $27\%$ away from the experimental value) 
and a much better prediction for $g_A=1.17$. Being the quantization of 
\cite{Amado:1986ef} equivalent to the standard one at large--$N_c$, we have no 
reasons to prefer, a priori, one or the other. We have no reason either, however, 
to believe that the $1/N_c$ corrections one includes in this alternative 
approach really capture the leading $1/N_c$ corrections or at least part 
of them. If this was the case we should, of course, use the 
non--standard quantization and the discrepancy in the prediction of 
$g_A$ would disappear.

If we stick, on the contrary, 
to the standard quantization procedure a small value of $g_A$ ($g_A=0.65$
\cite{Adkins:1983ya}) is also obtained in the original Skyrme
model, but the situation improves if the effects of the $\rho$ and $\omega$ mesons are taken into
account. The ``complete'' model described in Ref.~\cite{Meissner:1987ge}
seems the one which
should better mimic our 5D scenario, and $g_A=0.99$ in that case. The explicit
chiral symmetry breaking,
which is turned on in \cite{Meissner:1987ge}, could explain the difference because
the axial coupling is strongly sensitive to the large-$r$ behavior of
the solution (see the discussion following eq.~(\ref{axas}))
which is in turn heavily affected by the presence of the pion mass.
Correction to $g_A$ from chiral symmetry breaking could therefore be enhanced.
Notice that, however, this expectation fails in the original Skyrme model,
where the addition of the pion mass does not affect $g_A$ significantly \cite{Adkins:1983hy}.

In figs.~\ref{Fig:ElectricFormFactors}, \ref{Fig:MagneticFormFactors}
and \ref{Fig:AxialFormFactor} we compare the normalized nucleon form factors
at $q^2 \neq 0$ with the dipole fit of the experimental data.
The shape of the scalar and axial form factors
is of the dipole type, the discrepancy is mainly due to the error in the radii.
The shape of vector form factors is of course not of the dipole type for small
$q^2$, but this is due to the divergence of the derivative at $q^2=0$.
Including the pion mass will for sure improve the situation given that
it will render finite the slope at zero momentum; it would be interesting
to see if the dipole shape of these form factors is recovered in the presence of the pion mass.
 We also plot in the left panel of fig.~\ref{Fig:AxialFormFactor} the deviation of
ratio of the proton and neutron magnetic form factors from the large $N_c$
value which is given, due to the the different
large-$N_c$ scaling of the isoscalar and isovector components, by
$G_M^P(q)/G_M^N(q) = -1$. Not only
we find that this quantity is quite well predicted, with an error $\lesssim 15\%$,
but we also see that its shape, in agreement with observations, is nearly constant
away from $q^2=0$. Also in this case correction from the pion mass are expected to go
in the right direction.

\section*{Acknowledgments}

We would like to thank A.~Pomarol for the many useful discussions and for reading a preliminary
version of the paper. We also thank K.~Hashimoto, T.~Sakai, S.~Sugimoto for an interesting
e-mail correspondence and IFAE, where part of this work was done, for hospitality.
A.~W. thanks M.~Redi and R.~Rattazzi for discussions, the work of G.~P. was partially
supported by the European Union 6th framework program
MRTN-CT-2006-035863 "UniverseNet" and SFB-Transregio 33 ``The Dark Universe" by
Deutsche Forschungsgemeinschaft (DFG).

\appendix

\section{The Equations of Motion}

In this technical appendix we report the EOM for the 2D fields
which appear in our ansatz in eqs.~(\ref{sts}) and (\ref{ansk1}) and we explain
the notation used throughout the paper.

\subsubsection*{The Residual Gauge Invariance}

Before discussing the detailed form of the EOM,
it is useful to observe that our ansatz has not fixed the 5D gauge freedom completely, its form is indeed preserved
by chiral $SU(2)_{L,R}$ gauge transformations of the form
$g_R=U(t)\cdot g\cdot U^\dagger (t)$ and $g_L=U(t)\cdot g^\dagger\cdot U^\dagger (t)$ with
\be
g = \exp[i \alpha(r,z) x^a \sigma_a/(2r)]\,.
\label{eq:resU1SU2}
\ee
The operators $\Delta$ defined in eq.~(\ref{defde}) have simple transformation rules under the residual symmetry. Indeed
$$
\begin{array}{l}
g\Delta^{(1),ab}\sigma_a/2g^\dagger = \cos\alpha \Delta^{(1),ab}\sigma_a/2 + \sin\alpha \Delta^{(2),ab}\sigma_a/2\,,\\
g\Delta^{(2),ab}\sigma_a/2g^\dagger = \cos\alpha \Delta^{(2),ab}\sigma_a/2 - \sin\alpha \Delta^{(1),ab}\sigma_a/2\,,
\end{array}
$$
so that the 2D fields $\phi_{(x)}$ and $\chi_{(x)}$ defined respectively in eq.~(\ref{sts}) and (\ref{ansk1})
transform as charged complex scalars under this residual $U(1)$. It is not hard to see that the fields
$A_{\bar\mu}=\{A_1,A_2\}$ transform as gauge field, so that under a residual transformation one has
\be
\left\{
\begin{array}{l}
A_{\bar \mu} \rightarrow A_{\bar \mu} + \partial_{\bar \mu} \alpha(r,z)\,,\\
\rule{0pt}{1.25em}\phi \equiv \phi_1 + i \phi_2 \rightarrow e^{i \alpha(r,z)} \phi\,,\\
\rule{0pt}{1.25em}\chi \equiv \chi_1 + i \chi_2 \rightarrow e^{i \alpha(r,z)} \chi\,,
\end{array}
\right.
\ee
while all the other fields are invariant.

There is also a second residual $U(1)$ associated with chiral $U(1)_{L,R}$ 5D transformations of the form
${\widehat g}_R={\widehat g}$ and ${\widehat g}_L={\widehat g}^\dagger$ with
\be
{\widehat g} = \exp\left[i \beta(r,z)\frac{(k\cdot\x)}\alpha\right]\,.
\label{eq:resU1U1}
\ee
Under this second residual $U(1)$ only $B_{\bar \mu}=\{B_1,B_2\}$ and $\rho$ transform non trivially. We have
\be
\left\{
\begin{array}{l}
B_{\bar \mu} \rightarrow B_{\bar \mu} + \partial_{\bar \mu} \beta\,,\\
\rho \rightarrow \rho+\beta\,,
\end{array}
\right.
\ee
and therefore $B_{\bar \mu}$ is a gauge field and $\rho$ a Goldstone.

In order to make manifest the residual gauge invariance in the action and the
EOM we introduced gauge covariant derivatives for the $\phi$, $\chi$
and $\rho$ fields
\be
\left\{
\begin{array}{l}
\displaystyle
(D_{\bar \mu} \phi)_{(x)} = \partial_{\bar \mu} \phi_{(x)}
+ \epsilon^{(xy)} A_{\bar\mu} \phi_{(y)}\\
\displaystyle
(D_{\bar \mu} \chi)_{(x)} = \partial_{\bar \mu} \chi_{(x)}
+ \epsilon^{(xy)} A_{\bar\mu} \chi_{(y)}\\
\displaystyle
D_{\bar \mu} \rho = \partial_{\bar \mu } \rho - B_{\bar \mu}
\end{array}
\right.\,.
\ee

\subsubsection*{The Equations of Motion}

The easiest way to derive the EOM for the 2D fields is to
start from the Lagrangian and substitute the ansatz.
Using the 5D action in eqs.~(\ref{Sg}) and (\ref{Scs}) and rewriting the 5D
fields in terms of the 2D ones (eqs.~(\ref{sts}), (\ref{eq0m}), (\ref{ansk})
and (\ref{ansk1})), after a
straightforward computation one finds the expressions for the mass $M$
and moment of inertia $\lambda$ given in eqs.~(\ref{eq:mass}) and (\ref{eq:lambda}).
Notice that in order to obtain the order $K^2$ terms of the action
one has to perform a symmetric integration in $d^3 x$, which can simply be
implemented by the replacement $\x^i\x^j\rightarrow1/3\delta^{ij}$.
We report here two contraction identities of the ``doublet'' operators
$\Delta$ (eq.~(\ref{defde})) which can be useful for the computation of the
2D action
\bea
&\Delta^{(x),ab}\Delta^{(y),ac} = -\delta^{(xy)}\Delta^{(2),bc}+\epsilon^{(xy)}\Delta^{(1),bc}\,,\\
&\Delta^{(x),ab}\epsilon^{bid}\x_d = \epsilon^{(xy)}\Delta^{(y),ai}\,.
\eea

The EOM for the 2D fields can be simply obtained, at this point,
by imposing the variation of the 2D action to vanish. We have also checked the consistency
of our ansatz by showing that the same EOM are obtained by substituting
directly into the 5D equations (\ref{eomt}).
The EOM for the fields which are already turned on in the static case are
\be
\left\{
\begin{array}{l}
\displaystyle
\rule{0pt}{1.5em}D^{\bar\mu}\left(a(z) D_{\bar\mu} \phi\right)
+ \frac{a(z)}{r^2}\phi(1-|\phi|^2) + i\gamma L \epsilon^{\bar\mu \bar\nu}
\partial_{\bar\mu}\left(\frac{s}{r}\right)D_{\bar\nu}\phi = 0\\
\displaystyle
\rule{0pt}{1.5em}\partial^{\bar\mu}\left(r^2 a(z) A_{\bar\mu\bar\nu}\right)
- a(z) \left(i \phi^\dagger D_{\bar\nu} \phi + h.c.\right)
+ \gamma L \epsilon^{\bar\mu \bar\nu} \partial_{\bar\mu}\left(\frac{s}{r}\right)
(|\phi|^2-1)=0\\
\displaystyle
\rule{0pt}{1.5em}\partial_{\bar\mu}\left(a(z)\partial^{\bar\mu}s\right)
-\frac{\gamma L}{2 r}\epsilon^{\bar\mu\bar\nu}\left[
\partial_{\bar\mu}(-i \phi^\dagger D_{\bar\nu}\phi + h.c.) + A_{\bar\mu\bar\nu}\right]
= 0
\end{array}
\right.\,,
\ee
while the equations for the ``new'' fields which are turned on for the rotating skyrmion are
\be
\left\{
\begin{array}{l}
\displaystyle
\partial^{\bar\mu}(r^2 a(z) \partial_{\bar\mu} v) - 2 a(z)\left[
v(1+\left|\phi\right|^2) - \chi\phi^\dagger-\phi\chi^\dagger\right]
+\gamma L \epsilon^{\bar\mu \bar\nu}\left[
\frac{1}{2}(\left|\phi\right|^2 - 1)
B_{\bar\mu \bar\nu}+ r Q
A_{\bar\mu \bar\nu}\right]=0\\
\displaystyle
\rule{0pt}{1.5em}D^{\bar\mu}(r^2 a(z) D_{\bar\mu} \chi) + a(z)\left[2v\phi
-(1+\left|\phi\right|^2)\chi\right]
-\gamma L \epsilon^{\bar\mu \bar\nu}(D_{\bar\mu}\phi)
\left[i \partial_{\bar\nu}(r Q)+ D_{\bar\nu}\rho\right]=0\\
\displaystyle
\rule{0pt}{1.5em}\frac{1}{r}\partial^{\bar\mu}(r^2 a(z) \partial_{\bar\mu} Q)
-\frac{2}{r}a(z) Q\\
\displaystyle
\hspace{3em}-\frac{\gamma L}{2}\epsilon^{\bar\mu \bar\nu}\Big[
(i D_{\bar\mu}\phi (D_{\bar\nu}\chi)^\dagger + h.c.)
+ \frac{1}{2}A_{\bar\mu \bar\nu}(2 v - \chi\phi^\dagger-\phi\chi^\dagger)
- \frac{2}{\alpha^2}D_{\bar\mu}\rho\, \partial_{\bar\nu}\left(\frac{s}{r}\right)
\Big]=0\\
\displaystyle
\rule{0pt}{1.5em}\partial_{\bar \mu}(a(z) D_{\bar\mu}\rho)
-\frac{\gamma L}{2}\epsilon^{\bar\mu \bar\nu}\Big[
\left(D_{\bar\mu}\phi(D_{\bar\nu}\chi)^\dagger + h.c.\right)
+\frac{i}{2}A_{\bar\mu \bar\nu}(\phi\chi^\dagger - \chi \phi^\dagger)
+\frac{2}{\alpha^2} \partial_{\bar\mu}(r Q)\partial_{\bar\nu}\left(\frac{s}{r}\right)\Big]=0\\
\displaystyle
\rule{0pt}{1.5em}\partial^{\bar\nu}\left(r^2 a(z)B_{\bar\nu\bar\mu}\right)
+ 2 a(z) D_{\bar\mu} \rho\\
\displaystyle
\hspace{3em}+\gamma L \epsilon^{\bar\mu\bar\nu}
\Big\{\left[(\chi-v \phi)(D_{\bar\nu}\phi)^\dagger + h.c.\right]
+(1-|\phi|^2)\partial_{\bar\nu} v -\frac{2 r}{\alpha^2} Q\, \partial_{\bar\nu}\left(\frac{s}{r}\right)
\Big\} =0
\end{array}
\right.
\ee

In order to solve numerically the EOM, they must be rewritten as
a system of elliptic partial differential equations. This can be achieved by
choosing a 2D Lorentz gauge condition for the residual $U(1)$ gauge fields
\be
\partial^{\bar\mu} A_{\bar\mu}=0\,,
\qquad \quad
\partial^{\bar\mu} B_{\bar\mu}=0\,.
\label{eq:2Dgauge}
\ee
In this way the equations for $A_{\bar\nu}$ become
$J^{\bar\nu} = \partial_{\bar\mu}\left(r^2 a A^{\bar\mu\bar\nu}\right)
= r^2 a \partial_{\bar\mu}\partial^{\bar\mu} A^{\bar\nu}
+\partial_{\bar\mu}(r^2 a) A^{\bar\mu\bar\nu}$ which is an elliptic equation
and a similar result is obtained for $B_{\bar\mu}$.
As discussed in \cite{Pomarol:2007kr}, to impose the gauge condition, one can
solve the ``gauge-fixed'' EOM counting the gauge field components
as independent fields. In this way, if one imposes the gauge conditions
at the boundaries, then the gauge is maintained also in the bulk.

\subsubsection*{The Boundary Conditions}

The IR and UV boundary conditions on the 2D fields follow from
eq.~(\ref{irboundary condition}) and eq.~(\ref{uvboundary condition})
and from the gauge choice in eq.~(\ref{eq:2Dgauge}).
They are given explicitly by
\be
z=z_{\IR}\ :
\quad
\left\{
\begin{array}{l}
\phi_1 = 0\\
\partial_2 \phi_2 = 0\\
A_1 = 0\\
\partial_2 A_2 = 0\\
\partial_2 s = 0
\end{array}
\right.
\qquad\qquad
\left\{
\begin{array}{l}
\chi_1 = 0\\
\partial_2 \chi_2 = 0\\
\partial_2 v = 0\\
\partial_2 Q = 0
\end{array}
\right.
\qquad\qquad
\left\{
\begin{array}{l}
\rho = 0\\
B_1 = 0\\
\partial_2 B_2 = 0
\end{array}
\right.\,,
\label{eq:bcir}
\ee
and
\be
z=z_{\UV}\ :
\quad
\left\{
\begin{array}{l}
\phi_1 = 0\\
\phi_2 = -1\\
A_1 = 0\\
\partial_2 A_2 = 0\\
s=0
\end{array}
\right.
\qquad\qquad
\left\{
\begin{array}{l}
\chi_1 = 0\\
\chi_2 = -1\\
v = -1\\
Q = 0
\end{array}
\right.
\qquad\qquad
\left\{
\begin{array}{l}
\rho = 0\\
B_1 = 0\\
\partial_2 B_2 = 0
\end{array}
\right.\,.
\label{eq:bcuv}
\ee

The boundary conditions at $r=\infty$ have to ensure that the energy of the
solution is finite, this means that the fields should approach a pure-gauge
configuration. At the same time one has to require that the solution is non-trivial
and its topological charge (eq.~(\ref{Bch})) is non zero.
To obtain a soliton solution with $B=1$ one can impose the conditions
\be
r=\infty\ :
\quad
\left\{
\begin{array}{l}
\phi = -i e^{i \pi z/L}\\
\partial_1 A_1 = 0\\
A_2 = \frac{\pi}{L}\\
s=0
\end{array}
\right.
\quad\qquad
\left\{
\begin{array}{l}
\chi = i e^{i \pi z/L}\\
v = -1\\
Q = 0
\end{array}
\right.
\qquad\quad
\left\{
\begin{array}{l}
\rho = 0\\
\partial_1 B_1 = 0\\
B_2 = 0
\end{array}
\right.\,.
\label{eq:bcrinfty}
\ee

The $r=0$ boundary of our domain requires an ad hoc treatment, given that
the EOM become singular there. Of course this boundary is
not a true boundary of our 5D space, but it represents some internal points.
Thus we must require the 2D solution
to give rise to regular 5D vector fields at $r=0$ and we must also require
the gauge choice to be fulfilled. These conditions are
\be
r=0\ :
\quad
\left\{
\begin{array}{l}
\phi_1/r \rightarrow A_1\\
(1+\phi_2)/r \rightarrow 0\\
A_2 = 0\\
\partial_1 A_1 = 0\\
s=0
\end{array}
\right.
\quad\qquad
\left\{
\begin{array}{l}
\chi_1 = 0\\
\chi_2 = -v\\
\partial_1 \chi_2 = 0\\
Q = 0
\end{array}
\right.
\qquad\qquad
\left\{
\begin{array}{l}
\rho/r \rightarrow B_1\\
\partial_1 B_1 = 0\\
B_2 = 0
\end{array}
\right.\,.
\label{eq:bcr0}
\ee

\section{Numerical Techniques}

To obtain the numerical solution of the EOM we used the
COMSOL 3.4 package \cite{comsol}, which permits to solve a generic system
of differential elliptic equations by the finite elements method.
A nice feature of this software is that it allows us to extend the
domain up to boundaries where the EOM are singular
({\it i.e.} the $r=0$ line), because it does not use the bulk equations
on the boundaries, but, instead, it imposes the boundary conditions.

In order to improve the convergence of the program and the numerical accuracy,
one is forced to perform a coordinate and a field redefinition. The former
is needed to include the $r=\infty$ boundary in the domain in which the numerical
solution is computed. The advantage of this procedure is the fact that in this way
one can correctly enforce the right behaviour of the fields at infinity by
imposing the $r=\infty$ boundary conditions. A convenient coordinate change is
given by
\be
x = c \arctan\left(\frac{r}{c}\right)\,,
\ee
where $x$ is the new coordinate used in the program and $c$ is an arbitrary constant.
The domain in the $x$ direction is now reduced to the interval $[0,c \pi/2]$.
The parameter $c$ has been introduced to improve the numerical convergence of the
solution. A good choice for $c$ is $c\sim 10$, which allows to have
a reasonable domain for $x$ and, at the same time, does not compress the solution
towards $x=0$.

A field redefinition is needed to impose the regularity conditions at $r=0$
(eq.~(\ref{eq:bcr0})). For this purpose we use the rescaled fields
\be
\left\{
\begin{array}{l}
\phi_1 = x \psi_1\\
\phi_2 = -1 + x \psi_2\\
\rho = x \tau
\end{array}
\right.\,.
\ee
With these redefinitions, in the new coordinates, the $r=0$ boundary conditions
read as
\be
r=0\ :
\quad
\left\{
\begin{array}{l}
\psi_1 - A_1 = 0\\
\psi_2 = 0\\
A_2 = 0\\
\partial_x A_1 = 0
\end{array}
\right.
\quad\qquad
\left\{
\begin{array}{l}
\chi_1 = 0\\
\partial_x \chi_2 = 0\\
v = -\chi_2\\
Q = 0
\end{array}
\right.
\qquad\qquad
\left\{
\begin{array}{l}
\tau - B_1 = 0\\
\partial_x B_1 = 0\\
B_2 = 0
\end{array}
\right.\,.
\label{eq:bcr0new}
\ee

In order to ensure the convergence of the program another modification is needed.
As already discussed, to obtain a soliton solution with non-vanishing topological
charge we have to impose non-trivial boundary conditions for the 2D fields
at $r=\infty$ (eq.~(\ref{eq:bcrinfty})). It turns out that, if such conditions are imposed, 
the program is not able to reach a regular solution. This is so because the
$r=\infty$ boundary is singular and imposing
non-trivial (though gauge-equivalent to the trivial ones) boundary conditions
at a singular point spoils the regularity of the numerical solution; the same would
happen if the topological twist was located at $r=0$. To fix this problem we
have to perform a gauge transformation which reduces the $r=\infty$ conditions to
trivial ones and preserves the ones at $r=0$
 at the cost of introducing a ``twist'' on the UV boundary.
For this, we use a transformation of the residual $U(1)$ chiral
gauge symmetry associated to $SU(2)_{L,R}$ (eq.~(\ref{eq:resU1SU2})) with
\be
\alpha(r,z) = (1-z/L) f(r)\,,
\ee
where $f(r)$ can be an arbitrary function which respects the conditions
\be
\left\{
\begin{array}{l}
f(0) = 0\\
f(\infty) \rightarrow \pi
\end{array}
\right.
\qquad\text{and}\qquad
\left\{
\begin{array}{l}
f''(0) = 0\\
f''(\infty) \rightarrow 0
\end{array}
\right.\,.
\ee
For $c\sim 10$ it turns out that a good choice for $f(r)$ is $f(r) = 2 \arctan r$.
The gauge-fixing condition for $A_{\bar\mu}$ is now modified as
\be
\partial_r A_1 + \partial_z A_2 - (1-z/L) f''(r) = 0\,,
\label{eq:newgauge}
\ee
the UV boundary conditions are given by
\be
z=z_{\UV}\ :
\quad
\left\{
\begin{array}{l}
x \psi_1 = \sin f(r)\\
(-1 + x \psi_2) = -\cos f(r)\\
A_1 = f'(r)\\
\partial_z A_2 = 0\\
s=0
\end{array}
\right.
\quad\quad
\left\{
\begin{array}{l}
\chi_1 = -\sin f(r)\\
\chi_2 = \cos f(r)\\
v = -1\\
Q = 0
\end{array}
\right.
\quad\quad
\left\{
\begin{array}{l}
\tau = 0\\
B_1 = 0\\
\partial_z B_2 = 0
\end{array}
\right.\,,
\label{eq:bcuvnew}
\ee
and the $r=\infty$ constraints are now trivial
\be
r=\infty\ :
\quad
\left\{
\begin{array}{l}
\psi_1 = 0\\
(-1 + x \psi_2) = 1\\
\partial_x A_1 = 0\\
A_2 = 0\\
s=0
\end{array}
\right.
\qquad\qquad
\left\{
\begin{array}{l}
\chi = -i\\
v = -1\\
Q = 0
\end{array}
\right.
\qquad\qquad
\left\{
\begin{array}{l}
\tau = 0\\
\partial_x B_1 = 0\\
B_2 = 0
\end{array}
\right.\,,
\label{eq:bcrinftynew}
\ee
whereas the $r=0$ and the IR boundary conditions are left unchanged.
Notice that in the new gauge the EOM for $A_{\bar\mu}$ are modified
in accord to eq.~(\ref{eq:newgauge}), however they are still in the
form of elliptic equations.



\begin{thebibliography}{99}
%

\bibitem{Son:2003et}
  D.~T.~Son and M.~A.~Stephanov,
  Phys.\ Rev.\  D {\bf 69} (2004) 065020
  [arXiv:hep-ph/0304182].


\bibitem{Erlich:2005qh}
  J.~Erlich, E.~Katz, D.~T.~Son and M.~A.~Stephanov,
  Phys.\ Rev.\ Lett.\  {\bf 95}, 261602 (2005)
  [arXiv:hep-ph/0501128].

%
\bibitem{DaRold:2005zs}
  L.~Da Rold and A.~Pomarol,
  Nucl.\ Phys.\  B {\bf 721}, 79 (2005)
  [arXiv:hep-ph/0501218].

%
\bibitem{DaRold:2005vr}
  L.~Da Rold and A.~Pomarol,
   ``The scalar and pseudoscalar sector in a five-dimensional approach to
  JHEP {\bf 0601}, 157 (2006)
  [arXiv:hep-ph/0510268].

\bibitem{Skyrme:1961vq}
  T.~H.~R.~Skyrme,
  Proc.\ Roy.\ Soc.\ Lond.\  A {\bf 260} (1961) 127;   Nucl.\ Phys.\  {\bf 31} (1962) 556.

\bibitem{Adkins:1983ya}
  G.~S.~Adkins, C.~R.~Nappi and E.~Witten,
  Nucl.\ Phys.\  B {\bf 228}, 552 (1983).

\bibitem{Meissner:1987ge}
  U.~G.~Meissner,
  Phys.\ Rept.\  {\bf 161} (1988) 213.

\bibitem{Pomarol:2007kr}
  A.~Pomarol and A.~Wulzer,
  JHEP {\bf 0803} (2008) 051
  [arXiv:0712.3276 [hep-th]].

\bibitem{Pomarol:2008aa}
  A.~Pomarol and A.~Wulzer,
  arXiv:0807.0316 [hep-ph].

\bibitem{Sakai:2004cn}
  T.~Sakai and S.~Sugimoto,
  Prog.\ Theor.\ Phys.\  {\bf 113} (2005) 843
  [arXiv:hep-th/0412141]; {\it ibid.}
  {\bf 114} (2005) 1083
  [arXiv:hep-th/0507073].


\bibitem{Hata:2007mb}
  H.~Hata, T.~Sakai, S.~Sugimoto and S.~Yamato,
  arXiv:hep-th/0701280.

%
\bibitem{Nawa:2006gv}
  K.~Nawa, H.~Suganuma and T.~Kojo,
  Phys.\ Rev.\  D {\bf 75} (2007) 086003
  [arXiv:hep-th/0612187].


\bibitem{Hashimoto:2008zw}
  K.~Hashimoto, T.~Sakai and S.~Sugimoto,
  arXiv:0806.3122 [hep-th].


\bibitem{Hong:2007dq}
  D.~K.~Hong, M.~Rho, H.~U.~Yee and P.~Yi,
  Phys.\ Rev.\  D {\bf 77} (2008) 014030
  [arXiv:0710.4615 [hep-ph]].


\bibitem{Kim:2008pw}
  K.~Y.~Kim and I.~Zahed,
  JHEP {\bf 0809}, 007 (2008)
  [arXiv:0807.0033 [hep-th]].

\bibitem{Hata:2008xc}
  H.~Hata, M.~Murata and S.~Yamato,
  Phys.\ Rev.\  D {\bf 78} (2008) 086006
  [arXiv:0803.0180 [hep-th]].


\bibitem{Amado:1986ef}
  R.~D.~Amado, R.~Bijker and M.~Oka,
  Phys.\ Rev.\ Lett.\  {\bf 58}, 654 (1987).

%
\bibitem{Hirn:2005nr}
  J.~Hirn and V.~Sanz,
  JHEP {\bf 0512}, 030 (2005)
  [arXiv:hep-ph/0507049].



\bibitem{Finkelstein:1968hy}
  D.~Finkelstein and J.~Rubinstein,
  J.\ Math.\ Phys.\  {\bf 9} (1968) 1762.

\bibitem{Braaten:1986iw}
  E.~Braaten, S.~M.~Tse and C.~Willcox,
  Phys.\ Rev.\ Lett.\  {\bf 56} (1986) 2008.

\bibitem{Witten:1979kh}
  E.~Witten,
  Nucl.\ Phys.\  B {\bf 160} (1979) 57.

\bibitem{Witten:1983tx}
  E.~Witten,
  Nucl.\ Phys.\  B {\bf 223} (1983) 433.

\bibitem{Karl:1984cz}
  G.~Karl and J.~E.~Paton,
  Phys.\ Rev.\  D {\bf 30}, 238 (1984).

\bibitem{Beg:1973sc}
  M.~A.~B.~Beg and A.~Zepeda,
  Phys.\ Rev.\  D {\bf 6} (1972) 2912.

\bibitem{Gasser:1987rb}
  J.~Gasser, M.~E.~Sainio and A.~Svarc,
  Nucl.\ Phys.\  B {\bf 307} (1988) 779.

\bibitem{Dashen:1993as}
  R.~F.~Dashen and A.~V.~Manohar,
  Phys.\ Lett.\  B {\bf 315} (1993) 425
  [arXiv:hep-ph/9307241].

\bibitem{Adkins:1983hy}
  G.~S.~Adkins and C.~R.~Nappi,
  Nucl.\ Phys.\  B {\bf 233} (1984) 109.


\bibitem{Hong:2007kx}
  D.~K.~Hong, M.~Rho, H.~U.~Yee and P.~Yi,
  Phys.\ Rev.\  D {\bf 76} (2007) 061901
  [arXiv:hep-th/0701276].

\bibitem{Machleidt:1987hj}
For a review see
  R.~Machleidt, K.~Holinde and C.~Elster,
  Phys.\ Rept.\  {\bf 149} (1987) 1.

\bibitem{Amaldi:1972vf}
  E.~Amaldi {\it et al.},
  Phys.\ Lett.\  B {\bf 41} (1972) 216.

\bibitem{DelGuerra:1976uj}
  A.~Del Guerra {\it et al.},
  Nucl.\ Phys.\  B {\bf 107} (1976) 65.

\bibitem{comsol}
see http://www.comsol.com.

\end{thebibliography}
\end{document}